# Improving Interdisciplinary Communication with Standardized Cyber Security Terminology: A Literature Review

## RAMIREZ, ROBERT[1], CHOUCRI, NAZLI[2]

[1]Institute for Data, Systems, and Society, Massachusetts Institute of Technology, Cambridge, MA 02139, USA (email: roberto@csail.mit.edu)
[2]Department of Political Science, Massachusetts Institute of Technology, Cambridge, MA 02139, USA

**ABSTRACT** The growing demand for computer security and the cyberization trend are hallmarks of the 21st century. The rise in cyber-crime, digital currency, e-governance, and more, is well met by a corresponding recent jump in investment in new technology for securing computers around the globe. Recently, business and government sectors have begun to focus efforts on comprehensive cyber security solutions. With this growth has emerged the need for greater methods of collaboration and measurement of security. Despite all these efforts, this need has not been met, and there is still too little cross-disciplinary collaboration in the realm of computer security. This paper reviews the new trends in cyber security research, their contributions, and some identifiable limitations. We argue that these limitations are due largely to the absence of co-operation required to address a problem that is clearly multifaceted. We then identify a need for further standardization of terminology in computer security and propose guidelines for the global Internet multistakeholder community to consider when crafting such standards. We also assess the viability of some specific terms, including whether *cyber* should be used as a separate word when it is a descriptor (e.g. cyber security or cybersecurity), and conclude with recommendations for writing future papers on cyber security or the broader new field of all things relating to cyberspace, which has recently been dubbed *Cybermatics*, a term we also examine and propose alternatives to, like *Cyber or Cybernomics*. By furthering the effort of standardizing cyber security terminology, this paper lays groundwork for cross-disciplinary collaboration, agreement between technical and nontechnical stakeholders, and the drafting of universal Internet governance laws.

## I. INTRODUCTION

Cyber security is a nascent and exploding field with a growing body of research [77,78]. It is rooted in traditional computer science but has recently gained prevalence in other fields such as law and business management, as well as areas of technology that did not originally operate with the Internet, such as smart grids, cars, and other cyber-physical systems, which are experiencing new security vulnerabilities as a result of their new-found connections to it, although cyber-crime can be perpetrated without the Internet [112]. As a new field that sprang out of many old ones and serves as a unifying concern among disparate disciplines, cyber security has been given little attention in developing standards of research, possibly because these disciplines' standards have been specified strictly within their own field, or that the urgency of protecting against cyber-crime outweighs many standards, or perhaps that the field is still relatively new and many may think standardizing too soon could stifle growth – all of which are reasonable assumptions.

However, never has it been more urgent for cyber security to be unified as a well-defined and standardized academic discipline. Standardization is commonplace in scientific disciplines, beginning with either systematic nomenclature or otherwise standardized vocabulary [87-89]. Yet there have been very few efforts in research standardization, all of them government-led [118,119]. Herein we argue for and facilitate more formalized research by the global multistakeholder community, especially academia, in cyber security, beginning with facilitating greater communication among the disparate disciplines that concern themselves with this area, through identifying trends in terminology standards..

This paper first assesses the state of the field of cyber security with a literature review, covering many of its newer, less traditional aspects. The object of this study was to manually identify significant areas of focus in current academic cyber security research. Papers were selected manually from certain searches using the MIT libraries database. Given the extent of human ability, this manual portion of the study was not intended to construct any detailed ontology of cyber security. Inferences were drawn to identify many current trends, and papers were grouped by broad fields. These broad fields were loosely defined and posited to contain all research areas that were not explicitly identified. Some work on crafting ontologies of cyber security research has been done in the past using both manual and automated techniques [2, 30, 35, 58, 59, 62, 65, 74, 75]. However, such efforts usually used relatively few inputs, such as starting with a basic phrase of "cyber security" and performing automated searches for papers with this term; and therefore these studies may not have covered the entire scope of the field of cyber security.

To lay the foundations for standards in cyber security research, a unified terminology is essential. The majority of the body of this paper provides guidelines, metrics, and suggestions for unifying the terminology of cyber security research, using the myriad keywords taken from the literature review as a more "human-informed" basis for automated searches to identify trends. The incidence of the keywords as well as the incidence of all keywords from all papers with each of the keywords was recorded from searches on 2 major journal paper databases, Scopus, and IEEE Xplore. Trends were analyzed and, along with linguistic analyses and a test of trends of whether cyber is used as a prefix or as a separate word in journal papers from 1995 to 2005, recommended terminology standards criteria were established and baseline general standards were suggested. Some specific terms were also rigorously defined. This work is intended to serve as a guide to developing more standard terms or a more standardized terminology for cyber security by the academic community or governing bodies in the near future, or to be taken as guidelines for selection of keywords, titles, and proper technical terms in future cyber security research. The manuscript for this work was written to adhere to these terms except in the use of quotes from other papers, which, if conflicting, are indicated with [sic].

The specific contributions of this work are thus fivefold: 1. We identify a large proportion of the emerging trends in cyber security research; 2. We point out a long overdue need for standardization of cyber security terminology; 3. We propose guidelines to consider when selecting terms or standardizing terminology (e.g. prevalence and occurrence of terms, linguistics, governing bodies); 4. We propose and justify some actual terminology standards; 5. We identify the general silos cyber security research falls into, and their associated terminology, and suggest avenues for further research based on our classification.

## II. LITERATURE REVIEW METHOD

The intention of this literature review was to assess the state of emerging cyber security research and explore avenues of cyber security that have not received as much traditional attention as standard topics of network security, cryptography, and basic system security that a typical university curriculum in focuses on [105-107]. The manual literature review was performed via a number of particular searches throughout September 2015 with the

MIT libraries revolving around journal papers containing the word or prefix "cyber" and selected based on breadth of coverage as candidates for further reading. Selection criteria included priority given to other literature reviews and papers whose intention was broad characterization of issues, with a slight preference away from technical papers.

When technical papers were selected for review they were more often cyber-physical security papers, such as those on SCADA and PLC security. In addition, a number of papers from the Columbia University/Global Commission on Internet Governance (GCIG) 2015 Conference on Internet Governance and Cybersecurity [sic] were selected independently for review. In addition, the selection process evolved slightly over time, becoming more restrictive. The selection process for papers we identified using the MIT libraries online database is illustrated by the following search parameters, which were chosen to narrow down the majority of the papers selected for the literature review. Successive terms in each search (e.g. "review" followed by "overview") were added sequentially in time as the search was revised during the initial paper selection process in September 2015:

1. (cyber) AND (review OR overview OR meta-analysis OR survey OR primer OR literature OR outline OR governance OR international OR global OR sustainable) NOT (bullying OR psychology OR psychosocial)
2. cyberspace AND ( (review OR overview OR meta OR survey OR primer OR literature OR outline or sustainable) ) NOT ( (bullying OR psychology OR psychosocial) )
3. (cyber) AND (ontology)

All searches were restricted to academic journal papers or conference papers from the years 2012-2015. From these searches and the GCIG Conference, 134 candidate papers were selected. Further manual inspection for breadth of coverage was performed, with preference to topics less commonly covered in cyber security education. Time constraints for reading the papers also played a somewhat restrictive role in paper selection. 77 papers in total were selected from this group and were read or skimmed for content [1-74, 121-123]. The number of citations per paper was not known when selecting papers.

## III. RESULTS

After reading all of the selected papers, we analyzed them for commonalities. For the initial stage of the analysis, we grouped the selected papers into categories corresponding to the general area of research we concluded they concerned. Our proposed categories are outlined in Table 1 below. It was evident to us from the papers that these categories were appropriate because the authors of the selected papers typically wrote them in such a way that indicated that they were writing from a specific worldview of expertise, such as policy, or technology; or to a specific audience with such a worldview, rather than from an integrated worldview of cyber security that encompasses all of these areas. We qualitatively explore these categories in section 4.

| Table 1 | Categories of cyber security papers from the literature review | | | |
|---------|---------|---------|---------|---------|
| Label | General Descriptor* | General topics encompassed** | Extended High-Level Summary*** | References |
| A | Public Sector Policy | Global Internet law, politics, and governance | Arguably the most diverse category in this literature review. Contains papers on cyberspace geography and jurisdiction, papers on issues surrounding concepts of cyber war, Internet standards and governance, or various other legal issues, including crime problems that have a very legalistic or political angle. Papers containing broad questions of national security and strategy, the global multistakeholder community, and many other broad policy or international relations questions are included in this category. | 3, 7, 9, 12, 13, 14, 15, 20, 21, 23, 25, 26, 34, 36, 39, 41, 43, 45, 46, 49, 51, 52, 53, 54, 56, 64, 67, 69, 71, 121, 122, 123 |
| B | Cyberspace Infrastructure | Technical Cybersecurity Research (hardware, software, CPS, cryptography) | Most of the technological papers. While this literature review attempted to maintain a focus on cybersecurity of critical infrastructure, and thus that dominates this category, this category could theoretically include anything mathematical or related to software for the infrastructure of cyberspace. In our review, this includes analyzing SCADA vulnerabilities, high level papers on security of cyber-physical (e.g. critical infrastructure) systems, as long as they maintain some degree of technical rigor such as modeling. But it could also have included encryption scheme development, Internet protocol descriptions, and more. This category would not include software or models designed for policy analysis, business management, criminological or sociological papers, or ontologies aimed at broad descriptions of cybersecurity. | 5, 8, 16, 17, 18, 19, 24, 29, 38, 48, 55, 63, 66, 70 |
| C | Business and Operations | Business Management, frameworks, practices | These papers are by authors aiming to call attention to poor business practices and other kinds of risks for businesses, culture and awareness, and ways to improve these practices (many of which are simply negative statements: stop doing something, e.g. stop leaving CISOs/CIOs out of the loop). In addition, this category encompasses papers that include more in-depth *positive* frameworks for maintaining cyber security in a business through education, access control, supply-chain management, etc. That is, these frameworks specify what *to* do as opposed to what *not to* do. Many of these papers include not only suggestions for management but also propose research agendas for cyber security management. | 1, 27, 28, 31, 37, 40, 42, 44, 50, 57, 61, 68 |
| D | General Research | General cyber crime, threats, other broad analyses (including Ontologies) | These papers analyze what common cyber-crimes, attacks, and threats exist. These papers have a more technical slant than category A, but are much broader than B. In addition, this category contains ontologies for vulnerabilities and attacks, populating knowledge bases, or mapping links between threats and actors. Other inclusions in this category are such papers as "A systematic literature review of computer ethics issues" and other "system-wide" topics that are fundamental to understanding cyberspace. This category is typically aimed at a broader audience than the other categories. | 2, 4, 6, 10, 11, 22, 28, 30, 32, 33, 35, 47, 58, 59, 60, 62, 65, 73, 74 |

**Table 1.** Descriptions and summaries of proposed categories of cyber security research based on the literature review.
*Not meant to completely encapsulate the category out of context; simply meant as a means of easily identifying the general categories of journal papers we review herein. This is why we refer to the categories with the nebulous letters A–D. Categories may have some overlap and are intended primarily for organizing the topics found in our literature review. **In this review. *** See section 4 for illustrative summaries of select papers.

By identifying these categories, we hope researchers will consider them in the future when creating more formal categorizations or ontologies of cyber security. In this paper we merely call attention to our observations and do not claim to propose a formal structure delineating the boundaries of cyber security.

Identifying areas of cyber security research as we have done is an important step in formalizing the research methodology of cyber security, as it points to various fields to draw frameworks from, and helps frame research agendas. In section 6 we lay additional groundwork for this formalization for future researchers.

It became evident from our literature review and our identification of cyber security categories that there is a communication gap in cyber security dividing traditional technological research and the public and private sectors' nontechnical dealings with cyber security. This communication gap stands out to us as the largest fundamental problem impeding progress in this space by overlooking avenues for cross-disciplinary innovation.

## IV. PAPER SUMMARIES BY CATEGORY

Precisely defining each category is beyond the scope of this paper, as we do not wish to circumscribe research with more specific attempts at definitions. Instead, we describe the papers in more detail to illustrate some different emergent categories and subcategories of cyber security before analyzing the general shortcomings of the state of research in the field.

### A. PUBLIC

This category includes issues of concern to the government sphere of society. It includes work regarding norms, laws, and national security. Organizations such as ICANN and W3C, while often specifying norms, also create many technical standards. Such technology is not a part of this category, but we instead place it in the Infrastructure category. Also not in this category are topics like business operations and supply chain management, even if government services benefit from these topics. Instead, we discuss those topics in category C.

Harrop et al. (2013) give a short summary of cyber security efforts in the UK and the US, attempting to assess their protection measures, some of which address information sharing between entities on such topics as vulnerabilities and "cyber" incidents [25]. This includes a list of recommendations used by the UK Center for Protection of National Infrastructure (CPNI) to ensure security, and describes a number of UK efforts to help businesses and the nation address cyber security. They go on to describe the state of US cyber security such as the NIST cybersecurity [sic] framework [83]. They also list the critical national infrastructure sectors of the two countries.

Pawlak et al. (2013) analyze advances in threat evolution and government security, and compares them, concluding that governments need to do more to defend themselves and their states, starting with basic capacity building [49]. They say that nations are in danger of severely lagging behind trends in cyberspace. They also note, based on another study, eight innovations that will shape the future cyber security risk landscape: the cloud, big data, the internet of things, mobile internet, the neuronal interface, contactless payments, mobile robots, quantum computing, and the militarization of cyberspace. Lastly, they call for researchers to create a model of how exactly public and private spheres will collaborate in the future.

Grant et al. (2014) come up with cartographic terms for cyberspace and apply the concept of cyber-geography to military operations. They also suggest that research might be able to use their ontology to shed light on the attribution problem of being unable to expediently identify malicious actors through cyberspace [23].

Chertoff et al. (2015) describe the state of Internet jurisdiction law and the problem of assigning legal authority to a particular forum when a suit traverses multiple states. They propose four potential formulations that might clearly and fairly define the controlling jurisdiction in cases [9]. These formulations are choice-of-law rules based on either: the citizenship of the subject of the offending information, data, or system; the location where the harm has taken place; the citizenship of the data creator; or the citizenship of the data holder or custodian.

Lin (2015) compares nuclear and cyber technology and regulation, listing a host of differences, and a few similarities, between potential problems these two technologies create, which he places into categories of strategy, operations, acquisition, and arms control [26].

Common to all these papers, including the ones not mentioned here, is a notion of a system of governments that is lagging behind technology and that may not even be equipped to manage it well at all. They serve as a call to researchers and the global multistakeholder community alike to unite in search of solutions. They also point out a dangerous threat to governments and nations, not only in the form of cyber-attacks, but also in the form of other entities taking over to manage traditionally government-regulated matters, such as international communication, national security, and even control over borders and international law. These problems foreshadow many other problems to come for national governments, all of which are exacerbated by the existence of the Internet and widespread computing.

## B. INFRASTRUCTURE

This category includes most of the paper that address technological problems of cyber security, though more specifically, those problems related to the actual infrastructure of cyberspace, and not necessarily programming solutions to every problem that businesses or academic researchers might address – for example, a software tool for managing finances is a category C., Business, topic. The Infrastructure category includes papers that discuss various aspects of cyber security of critical infrastructure, as well as security issues concerning the operation of cyberspace, such as cryptography. It also encompasses papers describing methods for intrusion detection, reverse engineering, and computer forensics, among other issues.

Franke et al. (2014) systematically review 102 papers, drawn from IEEE Xplore, Scopus, Springer link, and Web of Science, in an effort to create a research agenda in the area of cyber situational awareness. Topics they cover include game theory, cognition, vulnerability detection, attack detection, other network analysis, broader primers; a great variety of articles on securing industrial control systems (ICSs)/SCADA (such as power grids); some concepts of emergency management; various tools, architectures, and algorithms on a host of topics, including attribution; and many papers on "visualization," for cyber situational awareness. They note deficits of papers in nation-wide or other high-level cyber situational awareness, despite cyber situational awareness being extremely popular with policy-makers; in teamwork (in various senses of the word) and information exchange, and in military strategy [16]. Franke et al. recommend more attention be paid to these areas, and also to efforts to deceive attackers, and to confidentiality and integrity. They suggest that researchers perform experiments to measure how particular solutions contribute to the overall understanding of a situation, and enumerate further directions for game-theoretic research, data fusion algorithms for low-level and high-level information, like sensors and NLP, and empirical work and exercises. Among efforts for cyber situational awareness, ICSs research is well-endowed.

Genge et al. (2015) provide a detailed description of their "cyber attack impact assessment methodology," which has the potential to be a general purpose tool to use in analyses assessing impacts from attacks on cyber-physical systems [18]. This carries implications for securing cities and countries with it, although these applications were not detailed in the paper.

Huang et al. (2015) detail an in-depth cyber-physical network architecture that, provably and in simulations, resists collapsing as a result of errors on either the cyber or the physical side, as a way of preventing cascading failures [29].

Gao et al. (2014) review a number of papers on SCADA implementation and security, providing a comprehensive reference. They describe two main categories of security issues in SCADA systems: direct threats (terrorist attacks, etc.) and indirect security threats (e.g. viruses, bugs). Gao et al. reiterate a common notion that SCADA security cannot be approached like traditional IT security, as availability and safety are paramount in SCADA systems, and SCADA infrastructure is less dynamic and less globally networked than traditional IT systems [17].

Cheminod et al. (2013) provide a literature review about the conceptual state of security for critical infrastructure and cyber-physical systems at the component and system level, including policy enforcement. They also provide a host of resources for industrial networks and mention future areas of research in great detail [8]. It is clear from these papers that there is no shortage of work being done on cyber-physical systems security. However, research of the connections and interactions of critical infrastructure with the rest of cyberspace and society is somewhat behind, as is research of the interaction of cyber-physical systems security and traditional cyber security. Not integrating cyber-physical security with concepts concerning other areas of cyber security is a common ailment among these papers. It is equally important for other STEM fields besides cyber-physical systems, who are now researching cyber security, to also integrate with traditional "computer science" cyber security researchers in this manner.

## C. PRIVATE

The third category we propose concerns business practices and other organizational and human factors affecting cyber security. The following papers give an illustrative overview of the types of papers in this category.

Messmer (2013) calls attention to the lack of coordination in businesses, which are also lagging behind technology [42]. She points to the fact that insurance decisions concerning cyber security are not discussed as often as they should be with C-level information officers. This problem is easily remedied, but it is a reflection of other organizational shortcomings in the workforce.

Khan et al. (2015) notes that the weak links in supply chains are often subject to attack. By analyzed the literature to identify if supply chain models can incorporate "cyber-resilience," they provided recommendations for practice as well as a number of research directions for identifying and securing against cyber-risk in supply chains [37]. "Cyber-resilience" is a popular buzzword that contrasts with cyber security by emphasizing the inevitability of cyber-attacks and the importance of being able to rebound as a business from such hiccups. However, proper cyber security education among employees and management is a better solution than implementing buzzwords and would eliminate the confusing notion that cyber security does not imply resilience.

Andel et al. (2013) surveyed various cyber programs at universities and retroactively document how a particular program developed at the University of South Alabama was created, detailing goals and objectives, and creating a curriculum that attempts to be comprehensive. They comment briefly on the problem of naming courses, which reflects the author's views on the necessity of a defined vocabulary: they give the example of Cybersecurity vs Cyber Engineering and the ambiguity of the differences between these two topics [1].

Sitnikova et al. (2014) write about a topic they say fall under "broader Internet security management and governance of the Internet and the cyberspace." They take a risk management approach to formulate a methodological framework for managing cyber security. They base their conclusions on a review of cases and previous studies, and highlight solutions at various levels of business operations, considering various elements of technology, people, and "processes," emphasizing that technology cannot solve all problems [57].

Jaitner et al. (2015) identify domains of science that contribute to the "cyber" field of study. They also identify points of necessary (presently implemented or not) collaboration across fields regarding a nation's cyber readiness. The goal of their paper is to identify areas not fully explored in academia, and for generating curricula recommendations. They aim to be comprehensive, drawing knowledge from Russia, and covering math, finance, linguistics, and natural sciences [31].

All the papers in this category point out that the organizational principles of businesses, and even fields of study, are not synchronized. There are vulnerabilities in supply chains, in trusting employees, and in social engineering. These papers also illustrate the point to be covered later about a lack of well-defined terminology and the prevalence of ad hoc phrases to describe certain, even redundant, aspects of cyber security. Moreover, papers in this category make clear that technological solutions alone cannot ensure cyber security.

## D. GENERAL

The "General" category contains all papers with issues which pervade the entire realm of cyber security, as well as descriptions of the field in general, and characterizations of cyberspace and humans' interactions with it.

Zhang et al. (2012) give a primer on, empirically, what actual crimes exist in cyber space. They categorize the crimes and call for action on existing problems such as cyber terrorism, phishing, and others [72].

Busse et al. (2015) give a helpful introduction to Ontology; specifically, contrasting the various meanings the word takes on in information science with social sciences. They conclude by stating

*"Different disciplines need to grow together more and more. The major challenges of our time – scientific and social – can only be solved interdisciplinarily. To be successful, it is vital that we manage to find results of various teams in various disciplines worldwide and to integrate them reasonably. Ontologies are of vital importance for this: by the power of standardizing terms, their meanings, and relations; furthermore, by the possibility of integrating different domain-ontologies; and, last but not least, by supporting the semantic web in search, reasoning and integration with computer applications. This is why we expect the importance of ontologies to grow significantly in future"* [6].

In short, Busse et al. provide a strong argument for cross-discipline communication and standardization of vocabulary terms.

Ju An et al. (2010) put forth a cyber security vulnerability ontology for comprehensive use, giving examples and references [35]. They are among many authors who propose information science-type ontologies, but do not necessarily scope the use of the ontologies, demonstrate their use, or make their ontologies publicly accessible.

Jardine (2015) gives an interesting perspective on the state of cyber-crime. He finds that most vulnerabilities are decreasing when normalized, and most attacks are increasing whether normalized or not, but are increasing more slowly over time and may soon be seen to be decreasing (essentially predicting a concave-down trend in attacks). In short, the picture of cyber-crime is not as bad as the absolute numbers make it seem when compared to the growth of cyberspace [33]. However, the data used may be imperfect and the trends are only analyzed from 2008-2014. His recommendations include: focus on the user rather than the system (i.e. put more effort into educating and empowering the user and continue putting the same effort into the technology of the system); use open-source code like SSL where possible to find vulnerabilities more quickly; create stricter rules for reasonable disclosure timeframes of zero-days by governments; develop international agreements on web-based attacks; create more cyber-crime insurance or other ways of spreading costs out; small-medium-sized companies need to invest in IT security and training as much as large companies; and cyber security companies should start to collect and represent their data in normalized terms.

Michael Chertoff, former Secretary of Homeland Security, in Chertoff et al. (2015) gives a primer on the dark web, the intentionally hidden part of the deep web, unindexed by search engines and impossible to reach with normal browsers. A traditional search engine sees about 0.03 percent of the web – the other 99.97% is the deep web. The authors assert that the global community needs to consider the deep web's impact when discussing Internet governance. A huge amount of crime (and a huge diversity of it) is supported in the deep web, and new ways to map and monitor it are needed. They nevertheless caution that the deep web's existence is good, in some ways, for everybody [10].

Common shortcomings of papers in this category are a lack of ontological understanding and scoping of the problems of cyber security, as well as indicators of a lack of a defined cyber security vocabulary across disciplines. These papers all conclude that the field needs more interdisciplinary cooperation, and that better characterization of cyber-crime and novel approaches to combating it, not necessarily technically, are imperative. Lack of consistent vocabulary is not in itself problematic – Busse et al. go into detail about the differences in the meaning of the word "ontology" between computer science, philosophy, and psychology – but the importance of such a paper is not to be understated [6]. To solve this problem in communication, the solution is itself communication. Standardizing vocabulary offers one outlet for such communication. Explaining differences in terms is another. Still another is creating businesses out of university research.

To summarize all four categories, it is evident from our literature review that there is a long-standing disconnect between traditional technological research in cyber security and the public and private sectors' nontechnical dealings with cyber security. This problem is likely a combined issue resulting from neither technical researchers nor management in business or government reaching out to communicate to the other parties. However, there is also a communication problem between researchers in the same category. Fundamentally, communication among researchers and between research and other sectors of society stands out as the largest general issue for cyber security when the field is broadly analyzed as we have done.

We might appear to take the supposed benefits of interdisciplinary research and communication for granted. Indeed, it may worry some readers that authors like Busse et al. propose such heavy collaboration between disciplines, that it might cause a kind of regression towards the mean if all disciplines standardized communication. However, that is not what we propose. We propose only that cyber security standardize some, if not all, of its terminology. Anything more is beyond the scope of this paper. Furthermore, we claim that not allowing disciplines to grow together (by not participating in this growth by utilizing concepts from different disciplines), is itself a regression towards the mean of one's own discipline, when innovation is at the edge – the edge of disciplines. While there are numerous edges to innovate on that research constantly takes advantage of, one particular edge – interdisciplinarity – is often overlooked. We postulate that interdisciplinarity and standardization create a new innovation edge, and we take this as the basis for our exploration of cyber security terminology standards that follows.

### E. COMPARISON TO PRIOR RESEARCH

Divisions of cyber security into four categories has recently been done by other initiatives as well. In 2015, the European CAMINO Project created the THOR acronym approach of "(T)echnical", "(H)uman", "(O) rganizational", and "(R)egulatory." The CAMINO Project asserts that cyber security can be comprehensively perceived as a combination of these four dimensions [82]. The THOR approach was put forth with the goal of creating an operational suggestion for a cyber security roadmap for Europe, and assumes integration of the four categories they proposed. This contrasts with our methodology of creating a classification of the state of research, which empirically highlights the lack of cooperation between the different categories.

### F. LITERATURE REVIEW CONCLUSIONS

One common ailment of all cyber security we determined from the literature review is that it is a poorly defined and new academic field subject to multiple and diverse definitions, with little educational basis, and few formalized research methods, especially outside of cryptography. Furthermore, most of its immediate implications lie outside of academia or the industry it caters to; it is a global and ubiquitous problem. Such a problem is difficult to formalize with research methods and education, to say the least. However, there are some operational measures that can be taken to improve the pace and quality of research; among them, is facilitating communication between scholars by standardizing terminology. It became apparent during the literature review that there is little to no standard terminology, especially outside technical cyber security, of which, cryptography is by far the most formalized, but even some of its practical implementations for private key encryption such as AES are supported not by rigorous mathematical proofs but by popular vetting [80].

Our focus herein is specifically the lack of consistency in nomenclature, such as, as Choucri, et al. wrote, whether to use "cyber" as a prefix, as in "cybersecurity" or as an adjectival modifier (i.e. a separate word, as in "cyber security" or "cyber-security") [75]. Sometimes even within the same article there is no displayed agreement on this convention, and authors may vacillate between the two [84].

## V. QUANTITATIVE ANALYSIS OF PAPERS
In the next sections of this paper we highlight the terms used in papers from our literature review, analyze the incidence of those and more general terms relating to cyber security, identify inconsistencies and their possible sources, and draw on linguistic and usage data to propose a basis for developing terminology standards for cyber security.

### A. METHOD
To analyze the meta-data of the papers reviewed, the author-supplied keywords of all the papers reviewed were extracted. In addition to author supplied keywords, some terms of interest were also extracted from titles. Scopus and IEEE Xplore were searched with dates from 2010 to 2015 to determine recent incidence of the terms. For added completeness, the following terms were added in addition: computer security, cyber domain, cyber war, cyber bullying, cyber physical, semantic web, semantic web search, cyber safety, cybernetics, sustainability, darknet, dark web, deep web, surveillance, cryptography, cryptology, encryption, and cryptanalysis. These searches were performed on Scopus c. 10/7/2015-10/13/2015 and on IEEE Xplore c. 10/12/2015-10/13/2015. The author-supplied keywords were searched for in quotes for exact terms (up to capitalization). Note that Scopus and IEEE Xplore treat hyphens as spaces. The number of hits for the terms were graphed on a logarithmic scale for ease of mental processing for both Scopus and IEEE Xplore. Both data sets and their sums are graphed in Table A1.

Table A1 categorizes the searched terms based on incidence with respect to powers of 10 in IEEE Xplore, Scopus, and in total. Double-counting in the total number of hits is not accounted for in this study, as only a general measure of academic use of these terms is sought from the data, and the hits from the individual databases (Scopus and IEEE Xplore) are considered in conjunction with the totals when conclusions are suggested herein.

Note that searches of IEEE Xplore returned about an order of magnitude less hits for most terms compared to Scopus.

### B. DISCUSSION OF DATA AND JUSTIFICATION FOR TERMINOLOGY STANDARDIZATION
Inspection of the terms in Table A1 reveals many that are undoubtedly unfamiliar or informal in appearance to most readers. The lack of consistent nomenclature observed in Table A1 is not limited to the question of how to form terms with "cyber," which itself is significant, but extends to whether "cyber" is always the most appropriate term, why we speak of e-commerce and not online commerce, why online psychology but cyber bullying or (recently) cyberpsychology [85].

While these terms may seem somewhat familiar to some readers, many other terms encountered when reading cyber security papers, especially nontechnical ones, are used rarely or only a few times, and can often seem ad hoc. Often when they are used, they are not rigorously defined or contrasted with other, perhaps more appropriate terms: should we speak of cyber security, cyber resilience, or cyber safety [79,82]? Many of these terms, given definitions of cyber security, are actually part of it: cyber security has been categorized into such stages as Identify, Protect, Detect, Respond, Recover, or with Confidentiality, Integrity, and Availability; and resilience and safety are arguably covered in those categorizations [83].

In some ways, use of cyber security terminology has begun to resemble the loosely-defined grammar of online fora, with contributors communicating in a mostly, but not always, mutually understood language that is just good enough [86]. With the rapid growth of cyber security, terminology standards that are just "good enough" will soon not be good enough, and may be even be overdue. Fields such as healthcare, chemistry, and electrical engineering all devote much effort to standards, including terminology [87-89]. While cyber security is not as old as these fields, losses from cyber-crime alone amount to approximately 1% of world GDP, and although this is not as large as health expenditures, it does not even reflect gains from ICT or the growing global dependence on computers [103-104].

A search of IEEE Xplore's standards dictionary returns only two records of standards terminology documents referring to "cyber" [90-92]. The lack of cyber security terminology standards is not only problematic in consistency of use, but in comparative studies and validity of results. Cryptography gives rigorous definitions of whether an encryption scheme is "secure," that allow schemes to be compared, but newer branches of cyber security, as well as broader "cyber" areas of study, are severely lacking such definitions.

Potential benefits of terminology standardization include the following:

- Creation of precise laws and policies
- Repeatable, mutually intelligible, and comparable research
- Preservation and availability of knowledge through easily searchable and indexed publications

Defining a term and eliminating unnecessary synonyms or ambiguous phrases from vocabulary facilitates the creation of precise legal constructs for cyberspace and the creation of industry standards and best practices, such as the NIST Cybersecurity [sic] Framework, which suggested standards for ensuring proper cyber security in business operations [83].

The ability for scholars to understand each other is of paramount importance in research and its vocabulary. As with the goals of chemical nomenclature, ensuring no ambiguity in terms should be of first importance, with a secondary objective being to minimize alternative names for the same concept.

This second objective would help database searches for new journal articles. If terminology constantly evolves, much knowledge can potentially be overlooked, with only the most common terms being searched for and recognized, and with papers using ad hoc or non-standard nomenclature being ignored or not even turning up in search results despite their valuable contributions. Therefore, to ensure accurate and comprehensive searches, standards benefit the entire body of research, and authors who choose not to adhere to such standards risk having a low impact on the field; therefore there is a strong incentive to adopt such standards if a plurality of researchers already have, and in many cases even if they have not yet done so [87].

## C. PRIOR RESEARCH STANDARDS WORK

Standardizing the field of cyber security has been an ongoing process for many years. In particular, there have been a number of attempts at the creation of a glossary of terms. The largest of efforts is the NICCS Glossary of Common Cybersecurity Terminology, a compilation of terms by US CERT from various lexicons issued by standards bodies [102]. These lexicons have been issued over the years by organizations like NIST. The East-West Institute has also led two smaller efforts in collaboration with the United States and Russian governments to create short agreed upon definitions of some terms, but these terms have been more specific to the defense sector [99-101].

Many of these cyber security lexicons seem themselves ad hoc or outdated, with terms like "misnamed files" and "mobile code;" and inspection reveals that many of the standards documents cited in them are over 10 years old [117]. Because its sources are old, NISTIR 7298 includes floppy disks and other removable media in its definition of "mobile devices," despite current usage of that term referring almost exclusively to smartphones. The field of cyber security is still new, but before 10 years ago it was in its infancy, especially from a government perspective, which most of the source documents cited in these dictionaries were generated from. Ten to fifteen years ago may have been too early to standardize cyber security terminology, at least without periodically updating it. CNSSI 4009, revised in 2006 and the most cited source used by NISTIR 7298 and the NICCS glossary, the two primary cyber security glossaries, states that a glossary must be continuously updated to remain useful and should keep pace with changes in cyber security [120]. While some of these have been updated over time, such as SP 800-53, many of these sources remain outdated.

Various terms from papers surveyed by our literature review do not appear in any of these dictionaries; terms like "big data", "cyber", "cyberbullying", "cyber-physical", "darknet", "internet of things", "smart grid", "web", and "Stuxnet", many of which in the past 10 years have become prominent, are notably missing from public sector definitions. Based on the contrasting terminology used in dated and government-defined dictionaries, we believe it is time that a coordinated effort between academia and industry, with input from government, took place, to update a comprehensive and representative cyber security dictionary of terms.

In addition to glossaries, other work related to research standards-setting includes a short and general research directive for allocating funds for cyber security research, The Cyber Security Research and Development Act (Nov 2002), which gave the US Office of Science, Technology, and Policy the responsibility for coordinating cyber security research and development. Besides this, there have been a number of cyber security research initiatives, largely supported by governments, but no broad industry or academia-wide efforts to create research standards; rather, these have been left to evolve organically [119]. A problem with this approach is that it took thousands of years for cryptography to evolve organically. Even concerted efforts have focused not on research standards, but security standards themselves, such as those for control systems, or for businesses [83, 118]. To our knowledge, a meta-level approach to cyber security has been largely neglected in research.

## VI. STANDARDIZATION RECOMMENDATIONS

In this section we will propose guidelines for authors and the global multistakeholder community in general to consider with coming to a consensus on standardized cyber security terminology. Here we develop standard terminology recommendation guidelines by analyzing the data in Table A1, and apply the guidelines to the keywords from the literature review.

### A. GUIDELINES

We propose the following guidelines for standardizing a cyber security term for a universal glossary, to reap the benefits stated above:

1. Clear linguistic basis as evidenced by etymology and adherence to proper rules of language.
2. Enjoys popular and historical usage by the global multistakeholder community based on trends in usage
3. Gives meaningful search results
4. Well defined and not ad hoc.

Herein we do not attempt to create new standards, but to posit inferred standards based on existing norms and inclinations of published works, in order to better facilitate research and discourse in the field. We hope to solidify emergent standards and avoid overburdening the research field with unintelligible phraseology.

We use these guidelines to present specific recommendations for terminology in section 6-C onward. In this paper, terms are not discounted for recommended standardization based on any one criteria. Throughout this paper, the following metric is used when suggesting standards: A term is recommended for standardization if it either: 1) explicitly satisfies at least 2 guidelines and does not explicitly fail to meet the other 2 guidelines, or 2) satisfies at least 3 guidelines.

While not all of these requirements may be *necessary* to recommend a particular term for standardization, by being strict in our selection of terms, we are guaranteed to satisfy more than *sufficient* criteria for acceptance. Future researchers may wish to more precisely incorporate dictionary terms to avoid the risk of overlooking important terms. We account for the concession by only seeking terms to accept, rather than terms to reject outright. Nevertheless, we note whether we accept or do not accept particular terms, in the following sections. *Acceptance* of a term means we recommend it for immediate standardization, whereas *Non-Acceptance* means we recommend it for sparing use in prominent places such as titles or keywords pending greater acceptance by the research and multistakeholder community. This paper is indifferent towards terms that are not explicitly commented on, with respect to whether we suggest them for inclusion in a cyber security lexicon at this time.

The above guidelines are for when there are no competing terms. If competing terms exist, the term that satisfies more guidelines is proposed; if they satisfy the same guidelines (such as in the case of two identical terms except one uses a "cyber" modifier and the other uses a "cyber" prefix), whichever one satisfies more guidelines to a greater extent (e.g. greater current incidence, earlier use or greater use over time, or greater acceptance by the global multistakeholder community) is given as our suggestion for the standard term.

Further elaboration on the measurement criteria for each of the proposed guidelines follows below.

1. The Elements of Style and various linguistic accounts, including journal publications and use by country and in government documents, will offer insight into etymology and proper English use [97].
2. Trends of usage over time from Scopus and IEEE Xplore will provide evidence of historical acceptance. To a lesser extent, use by agencies and working groups in the global multistakeholder community will also be examined for consistency with results from databases. Because the primary goal of this paper is to propose nomenclature for *research*, not for individual working groups or agencies for internal use, the primary sources will be results returned from academic journal databases. For this guideline we determine when terms first began to enjoy use among researchers.
3. Meaningful search results for journal database searches will be determined by returning hits in a range of incidence which is not too high nor too low, that we establish below. This range is empirically derived based on the incidence of currently accepted terms or candidate terms, such as the incidence range of "cryptography" or "cyberspace". The aim of defining such a range is to include all relevant terms, while excluding ad hoc terms and terms that are too broad to be meaningful outside of more specific contexts, such as "information." This ideal range will vary between databases, but will itself be used in this paper to include or exclude terms, which is the primary goal of this section. By adhering to data from a particular set of databases, we can identify appropriate terms. Because the ideal range will vary, it should only be used by other researchers 1) on the same databases, 2) within the same range of years (2010-2015).
4. The presence of rigorous definitions in journal articles is required to satisfy this guideline. Even for a

popular term, this requirement might not be satisfied. This is also measured by the number of overlapping or conflicting terms, e.g. online psychology versus cyberpsychology, the latter of which is not easily understandable. Definitions (extracted from dictionaries and journal articles) go beyond proposing words for broad concepts, and rigorously define these terms. For example, terrorism is a well-accepted term, but the definition of terrorism is highly contested [93].

### B. MEANINGFUL SEARCH RESULTS

We claim, as shown in Guideline 3, that searchability is an important guideline to consider when agreeing on a standard dictionary of academic terms. By searchability, we mean that performing a particular search (that is, with a particular keyword or phrase) gives meaningful search results, corresponding to papers the searcher was looking for. That is, the intended meaning of the keywords in the paper corresponds to the use of that keyword in papers in the database. The more appropriate the author-selected keywords, the more likely the author's paper is to appear in an appropriate search. We take this to be a primary quantitative identifier of whether a term should be a candidate for standardization. Identifying inappropriate search terms, while in itself does not exclude such terms as candidates for a standard cyber security vocabulary, does by itself provide guidelines for increasing visibility when choosing title, abstract, and keywords. We believe the optimal range where candidate standard terms can be found is [100,1000) total hits in Table A1, whereas the minimum range for candidate terms is [10, 100000). The following paragraphs elaborate on this claim.

In Table A1, terms in category 1 are clearly poor search terms. They have no value as keywords because of their gross ambiguity and universality. They are recommended to never be used as keywords. Taken as a whole, the terms in category 2 from Scopus give a broad idea of concepts in cyber security, but individually, these terms can have many meanings independent of cyber security; take space, ecosystem, sustainable, and planning, for instance. Even "internet" and "security" are a little too broad for our purposes of identifying a minimum vocabulary; minimum meaning with the strictest inclusion criteria to ensure that all terms selected unequivocally belong to cyber security and would turn up in a reasonable search for publications, and not in searches in vastly different fields like biology.

Scopus category 3 contains some words like cyber, encryption, network security, and smart grid, which clearly belong in the field and would make for search terms which only return appropriate publications. However, terms like geography, supply chain, and ontology have many applications to other fields, which makes them terms unlikely to return useful results on their own. Moreover, researchers should not be expected to sift through 10,000 of more papers to find relevant ones, unless perhaps they intend to do a broad literature review, such as the one in this paper. Therefore, while terms in Scopus category 3 highlight key high level aspects of cyber security, like cryptography, in actuality, a search for "cryptography" by itself will not yield anything specific enough to be of value without performing more in depth analysis of the search results. Therefore, this range of incidence is not specific enough on its own to be of value for Scopus searches. The above conclusions similarly apply to IEEE Xplore's categories 1-4. Category 3's upper bound of 100,000 is hits is only recommended as the upper bound for the least specific keywords used in an article.

Scopus category 4 terms are nearly all unambiguously and readily identifiable as specific to cyber security. They are still broad within cyber security, and are more appropriate as standalone search terms for specific literature reviews within cyber security. However, they do give meaningful search results. Terms in this range in Table 1 for Scopus and ranges of terms with similar incidences in Scopus in other databases may be an appropriate upper bound for inclusion as keywords, but because this paper aims to recommend concrete guidelines to describe a minimum number of appropriate cyber security search terms, category 4's range is still too high for this purpose. These terms would, however, be expected to yield specific meaningful results in searches when combined with other terms.

Every term in Scopus category 5, with the exception of perhaps "index terms" is clearly a cyber security-relevant term. Furthermore, the low number of hits in search results of these terms is manageable for anyone to sort through to identify papers of interest. This range's upper bound of 1000 is recommended as the upper bound for the most specific keywords used in journal papers.

Scopus category 6 contains a number of terms like cyber law, cyber insurance, and hacktivist that, while many may argue are valid vocabulary for inclusion as standards in the cyber security lexicon, do not yield very many search results, and furthermore, are not universally accepted or distinguishable from other aspects of cyber security. Cyber conflict is not easily distinguishable from cyber war, and the advantage of using terms like "safety" and "resilience" in place of security is not justified by papers using them [95,96,99]. Furthermore, many readers may find some of these terms unfamiliar. Given this, while category 6 may outline areas where further research is needed, to maximize visibility and yield results in meaningful searches, category 6 is not recommended except perhaps as the lower bound for the most specific terms used as keywords.

Category 7 needs no discussion given the above; it is peppered with ad hoc terms of little value, as evidenced by their low incidence. It may be a useful reference to identify future research directions, but it is not recommended that any terms in this category ever be used as paper keywords.

In summary, the optimal range where candidate standard terms can be found is [100,1000), whereas the minimum range for candidate terms is [10, 100000). Searches using terms in this latter range that do not fall in the former range should be performed by combining multiple terms to yield the most meaningful search results. When suggesting standard terms and rejecting others, we considered the optimal range of [100,1000) total hits when assessing whether guideline 3 was satisfied by a given term (see Table 2 below). While using sub-optimal terms and phrases may be an indispensable aspect of the progression of research, to ensure that publications are locatable, we would recommend that at least some of the author-supplied keywords be ones that are more easily searchable; querying databases with target journals can aid authors in this decision.

### C. RECOMMENDATIONS FOR SPECIFIC TERMS

The keywords extracted from the papers found in the literature review, as well as a dozen other terms we believe to be important in cyber security, are categorized in Table 2 based on our recommendations for standard usage. All terms were evaluated using the terminology guidelines outlined in this paper, and were then sorted into three categories, of either *Accepted*, *Not (yet) Accepted*, or *Partially Accepted*, based on the degree of their adoption by researchers and other members of the global multistakeholder community, as determined by the number of guidelines they satisfied. As stated before, terms that 1) explicitly satisfied at least 2 guidelines and did not explicitly fail to meet the other 2 guidelines, or 2) satisfied at least 3 guidelines were classified *Accepted*.

We make no we recommendation that terms we found not to be commonly accepted not be used. Table 1 only labels words according to their use in cyber security. Some words that are not yet accepted include "cyber" by itself (and in its myriad ad hoc combinations) and "cybernetics", as well as "cyber-risk" and "ontology". Although some of such "not accepted" terms may be understood by the reader, and may be well-defined in other fields, these terms are not yet generally understood within most of the cyber security academic and multistakeholder community. We do advise to not include such terms in current glossary updates, until they become more universally accepted and identified with cyber security.

Partially accepted terms in Table 1 are recommended to be prominently used in papers, such as in the title or author-supplied keywords, only occasionally, with discretion. For example, while "risk" may be an inappropriate author-supplied keyword, it is an acceptable term for use when describing topics in cyber security elsewhere in a paper. These "partially accepted" terms only satisfied one of our proposed guidelines, without outright failing to meet the other three, or met two guidelines but failed the other two. For example, according to Figure A1a, "critical infrastructure" is orders of magnitude more popular than "critical national infrastructure." Furthermore, the US CERT Cyber Glossary only defines critical infrastructure, not critical national infrastructure [102]. Therefore, we recommend that "critical infrastructure" be used and "critical national infrastructure" or CNI not be widely used at this time. Of course, CNI may still very possibly become a standard dictionary term in the future.

### VII. SPECIFIC NOMENCLATURE

In this section we elaborate on some of the more prominent cyber security terms categorized in Table 2, and their associated hindrances to the creation of a standard glossary of terms. Here we resolve some longstanding confusion and determine appropriate usage of some important terms.

### A. CYBER AS A MODIFER: ONE OR TWO WORDS?

To resolve the conflict of whether terms should use "cyber-" as a separate word (with or without a hyphen) as in "cyber attack" or "cyber-crime" or rather as a prefix of a word as in "cyberspace," the historical incidence of terms containing "cyber" was determined and linguistic analyses were performed.

First, IEEE Xplore was searched for articles containing "cyber" only as a word and those containing "cyber*" as either a word or as part of a word, where the asterisk indicates a wild card. The difference between the two terms was taken to yield only cyber* as a prefix/part of a word. The usage of "cyber-" as a word and of "cyber*" as a prefix were plotted from 1990 to 2015 after being controlled for the occurrence of "cybernetics." (Figure 2) This was done to ensure that only terms relevant to cyber security or the broader "cyber" research field were accounted for.

The majority of the words that appear after "cyber" (with a space or hyphen) in journal papers come from cyber physical, cyber security, cyber attack, cyber threats, cyber crime, cyber warfare, cyber world, and cyber war. Google Ngram also indicates other commons terms like cyber space [sic] [94]. The top terms containing "cyber" as determined by Google Ngram and Figure A1 were also plotted between 1990 and 2015 in Figure 2; and to control for such more common terms that dominate some of the "cyber" categories, like cyberphysical and cyberspace, curves that also control for these terms were plotted as well. These curves are bolded and labeled as "controlled." This was done in order to compare whether "other" generic terms, including ad hoc terms and terms that are simply less common, were more commonly used with "cyber" as a separate word or as a compound word. That is, whether "cyber" as a word or "cyber" as part of a compound is more commonly used in research articles.

| Table 2 | Summary of Proposed Terminology Standards | | | |
| Accepted | Not (yet) accepted | | | Partially accepted |
| --- | --- | --- | --- | --- |
| CISO | academia | hierarchical access | research strategy | accountability |
| cloud computing | active air defense | impact assessment | risk assessment | attack |
| computer abuse | active air defense | index terms | scientific paper | availability |
| critical infrastructure | active cyber defense | information exchange | secure software engineering | big data |
| cryptanalysis | adaptation tactics | information extraction | security analysis and monitoring | cascading failure |
| cryptography | ami | information schema | security automation | cio |
| cryptology | attack description language | information security educa- | security countermeasures | cni |
| cyber crime | | tion | security issues | computer crime |
| cyber law | attribution | information structure | security methodologies | Common vulnerabilities and |
| cyber operations | cikr | insider | security ontology | exposures |
| cyber physical | classification | instrumental crimes | security solution frames | computer ethics |
| cyber physical systems | communication | international | self-organisation | Computer security |
| cyber security | complex networks | international cooperation | risk assessment | computer system security |
| cyber threat | computational part | international policy | scientific paper | context-awareness |
| cyber war | cpss | internet security | secure software engineering | cps |
| cyber warfare | cross-domain attacks | internet study | security analysis and monitoring | cyber bullying |
| cyberspace | curriculum development | jurisdiction | security automation | cyber insurance |
| darknet | cyber | knowledge base | security countermeasures | cyber stalking |
| DDOS | cyber assurance | knowledge model | security issues | Cybercrime |
| deep web | cyber attacks and countermeas- | law | security methodologies | Cybersecurity |
| denial of service | ures | layered network | security ontology | dark web |
| digital signature | cyber conflict | learning objects | security solution frames | darknet |
| embedded computer | cyber domain | legal issues | self-organisation | e-commerce law |
| encryption | cyber education | legal rights | semantic | evidentiary |
| espionage | cyber psychology | literature review | semantic operability | forensics |
| hacker | cyber readiness | mac security | semantic security | hacktivist/hactivist |
| ict | cyber resilience | mapping | semantic web search | industrial networks |
| ids | cyber safety | meta-adaptation strategies | semantic web technology | information |
| information technology | cyber space | military operations | Slovenia | information systems security |
| internet | cyber targeting | model-based design | social cybernetics | insider |
| intrusion detection system | cyber treaty | morality of law | sovereignty | missile defense |
| malware | cyber world | socialization | space | risk assessment |
| national security | cybergeography | motivation | state-level | risks |
| network security | cybernetics | multi-agent systems | sustainability | security |
| phishing | cyber-physical-social systems | national cyber strategies | sustainable | security architecture |
| privacy | cyber-risk | networked computer tech- | system dynamics | security controls |
| risk management | cybers | nology | systematic literature review | security patterns |
| scada | cybersafety | neutralization | system-level requirements | self-defense |
| steganography | cyberspace security | ontology | systems strategic security man- | semantic web |
| stuxnet | cyber-territory | ontology architecture | agement | sensitivity analysis |
| system security | definitional gaps | ontology design | taxonomy | signals intelligence |
| | denial of sustainability | ontology security | technology | situational awareness |
| | deterrence | ontology-based context | Terrorism | smart grid |
| | disgruntlement | models | textbook | social-networking |
| | distributed systems security | organizational justice | theoretical foundation | software piracy |
| | e-consumer protection | papa framework | traceability | supply chain |
| | ecosystem | people | u.n. | supply chain management |
| | emerging cyber threats | percolation theory | us cyber security act 2012 | threat |
| | emerging technology trends | physically-aware engi- | vishing | threat environment |
| | employee computer crime | neered systems | web attacks | threat patterns |
| | ethical issues | planning | web space | u.s. cyber command |
| | expressive crimes | policy | propaganda | vulnerability analysis |
| | force | policy making | psycho dynamism | web |
| | geography | politics | regulation | |
| | government response | private sector | | |

**Table 2.** Keywords extracted from our literature review, and additional cyber security terms, grouped according to whether they satisfy section 6's guidelines.

**Figure 1.** Incidence of the most commonly appearing terms with the word *cyber* in journal papers, from Scopus.

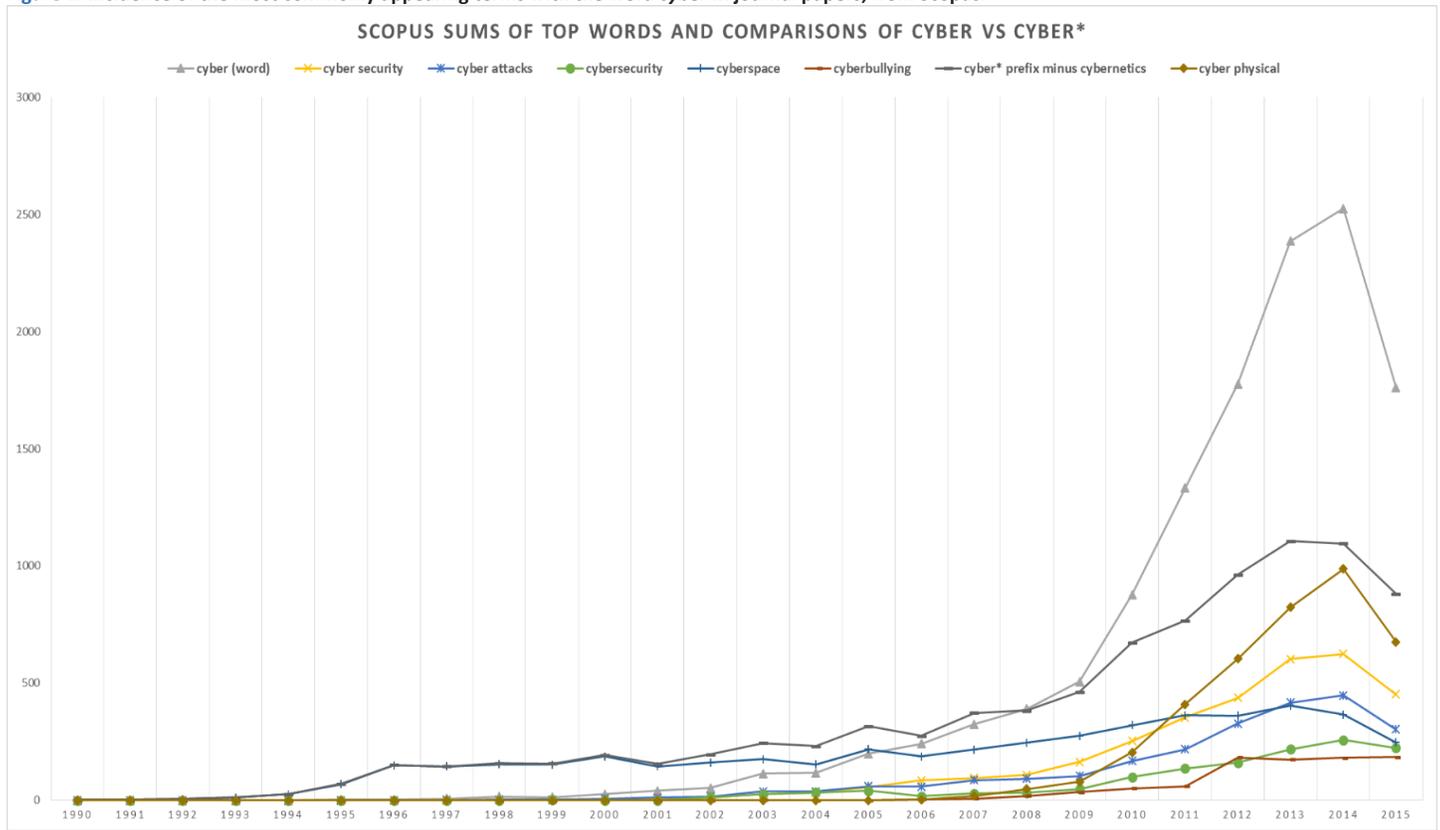

**Figure 2.** Incidence of the most commonly appearing terms with the word *cyber* in journal papers, from IEEE Xplore.

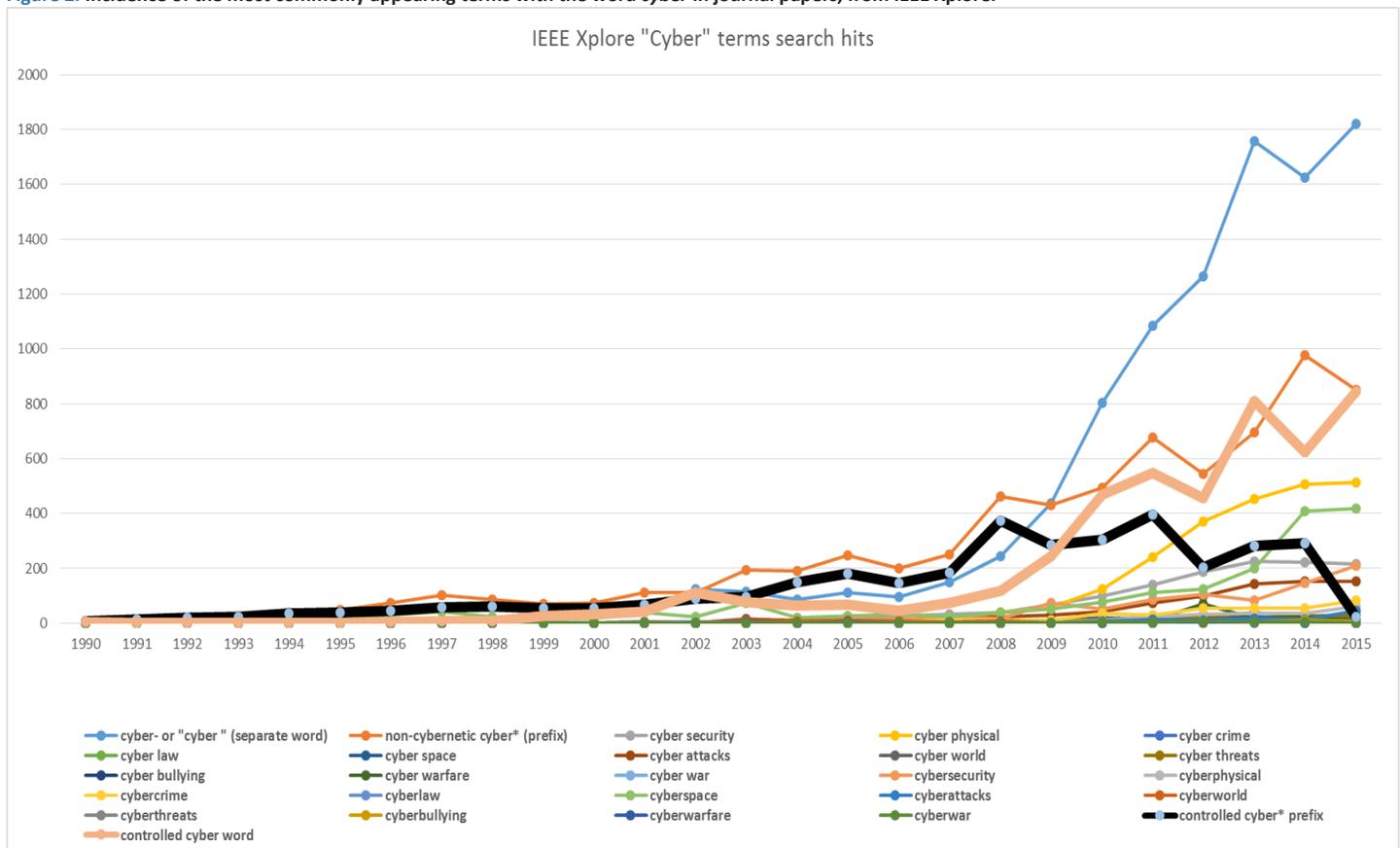

Similarly, Scopus was queried for the most common terms using "cyber." However, Scopus does not have a wildcard search as of this writing, so it is not possible to extract the exact number of terms that use "cyber" in a compound. However, summing the hits for the most commonly used "cyber" terms (other than cybernetics) for the two types yields an approximation of the totals of the two types. These approximations, along with the hits of some of the most common terms, were plotted (Figure 1).

The results indicate that both when controlling and when not controlling for the most commonly used terms containing "cyber," use of a separate word for "cyber-" is vastly more common than use in a compound word as of 2009; whereas prior to 2009, both had comparable incidence. Therefore, it is recommended that "cyber-" be used in most cases. In Figure 1, the controlled "cyber-" word is even beginning to overtake the incidence of all (non-cybernetic) compound words, whereas the use of compound words outside of the few most common ones is not gaining additional acceptance by the academic community. Figure 2 tells a similar story. "Cyber-physical" is by far the most prevalent term with "cyber" as a separate word, threatening to overtake the incidence of "cyber" in compounds. The separate word far outstrips the compound in total non-cybernetic hits.

It is clear from Figure 1 and Figure 2 that "cyber-" as a separate word, possibly hyphenated (according to preference), should be the standard format to ensure searchability. In database search engines that do not allow wildcard searches for word prefixes, having "cyber" as a separate word is extremely valuable, as it allows new and unfamiliar terms to be discovered. If a single compound word is used to search, only by already knowing the exact word one is searching for (which is unlikely given the nascence of the field) can appropriate articles be located.

This conclusion is consistent with guidelines 2-4 above for standardizing terms, but not with the 1st guideline. However, it is far better than the alternative, which only satisfies the first guideline. Historically, as can be seen from Figures 1 and 2, "cyber" as a separate word enjoyed less usage than similar compound words. Only around 2008 did it overtake the historical word; however, the separate word's usage so vastly outpaced the compound word's, that it is impossible to resist its current prevalence. Both standards that were returned from a search of IEEE Xplore's standards dictionary were of "cyber security," not "cybersecurity;" one was from 1997, and the other was from 2010 [90-92].

Finally, the linguistics of *cyber* should be considered to give it proper treatment under Guideline 1. The Greek root *κυβερνήτης* is not a compound word, and cyberspace and cyberwar can be thought of as portmanteaus of cybernetics and space and war, respectively [128]. Portmanteaus are nearly always single words, not containing hyphenated word fragments or word fragments separated by a space. However, unlike many portmanteaus, the second word is present in its entirety in both of these examples. Alternatively, since cyber is a standalone word that originated as an abbreviation of cybernetics, it might make more sense for it to appear as a separate word in compounds, especially when the full word it modifies is retained. In addition, while UK and European English sometimes appear to favor "cyber security" over "cybersecurity" (often favored by the US government), regional preferences have blurred recently. The ambiguous linguistic status of *cyber* is almost enough for Guideline 1 to yield little guidance, but we believe that the etymology of the word *cyber* favors a separate word usage in most forms.

Despite these observations, there are terms that are commonly used in a compound form. Among these are "cyberspace" and "cybersecurity." Curiously, cyber security and cybersecurity have comparable incidences in all of our figures that they appear in, although the separated-word phrase is still used about twice as often as the compound word.

### B. CYBERSPACE

"Cyberspace" actually meets all of the proposed guidelines for standardization, (except, arguably, the 1st guideline), and should therefore continue to be used frequently. "Cyberspace" emerged in 1990 according to Scopus, enjoys popular use, gives meaningful search results (see Table A1), and is consistently favored over "cyber space". Therefore, we suggest "cyberspace" as an appropriate standard term, and suggest "cyber space" never be used.

## C. CYBERSECURITY VERSUS "CYBER SECURITY"

"Cybersecurity" meets guidelines 3 and 4, but among journal papers, does not enjoy pluralistic usage over "cyber security," and in fact emerged after "cyber security," which has enjoyed more popular usage than "cybersecurity" in nearly every year according to both Scopus and IEEE Xplore's databases. In addition to this, while two words usually become one after a period of hyphenation (or a space; journal databases treat hyphenated words as separate words), the research community does not seem ready to accept cybersecurity as a single word yet [97]. However, due to their reasonably comparable incidences over time, "cyber security" or "cybersecurity" are both common and generally acceptable. However, this requires that searches for papers referring to cyber security include "cyber security" OR "cybersecurity" for complete coverage. This is of course tedious, and at this stage "cyber-security" or "cyber security" is recommended as the standard term over "cybersecurity" because it satisfies all four guidelines, whereas "cybersecurity" only clearly satisfies 2, 3, and 4, and satisfies 2 to a lesser extent than "cyber security" does.

## D. CRYPTOGRAPHY, CRYPTOLOGY, CRYPTANALYSIS

Cryptography refers to the art of designing cryptosystems, cryptanalysis refers to the art of breaking cryptosystems, and cryptology is the union of cryptography and cryptanalysis [98]. However, "cryptography" and "cryptology" are sometimes used interchangeably, although these terms are fairly well-defined in principle. In practice, "cryptography" is used far more widely than either "cryptanalysis" or "cryptology" according to Figure A1a and Figure A1b. Clearly this simply means that in the field of cryptology, significantly more effort has been devoted to cryptography than to cryptanalysis or to discussions of the general field. Both terms satisfy all four guidelines for standardization.

Given the definition of the words, we recommend that cryptology and cryptography be used properly in the future. However, there is, by definition, overlap in the two terms, so in cases of overlap, the more specific term, which is also the more prevalent term, "cryptography" should be used.

"Cryptanalysis" seems to have fallen into disuse and should therefore not be used as a primary search term when cryptography is a better alternative, given that, by definition, cryptanalysis seeks to break the cryptosystems of cryptography; that is, cryptography is implied in cryptanalysis, but not vice versa. Since "cryptography" is the most exclusive, or most essential of these three terms, it is recommended that this trend in usage be followed by authors to ensure visibility of publications. Again, this is not to say that cryptanalysis is a poor word choice. This paper makes no claims about the usefulness of words, only suggestions for which terms can be readily turned into universal standards.

## E. CYBERCRIME AND COMPUTER CRIME

According to Scopus, "computer crime" first appeared in the literature in 1972, well before either spelling of cyber-crime. Therefore, *computer crime* satisfies guideline 1 and *cybercrime* fails at guideline 1. However, many organizations in the global multistakeholder community refer to *cybercrime*, including Norton, Interpol, and the US government, though some do refer to *cyber-crime*, especially non-US countries [113, 114]. *Cyber-crime* as a hyphenated word appeared in the literature a few years before cybercrime in the mid-1990s, but *cybercrime* has in recent years begun to outpace *cyber-crime* in journal article usage. While both *cybercrime* and *cyber-crime* fall in the idea incidence range of [100,1000] hits from 2010-2015 on Scopus and IEEE Xplore combined, computer crime has far more hits, with 11,171 from Scopus alone. Although this is outside the ideal range, it is within the acceptable range. Thus *cybercrime* meets guideline 3 for meaningful search results, and both *computer crime* and *cyber-crime* satisfy guideline 3 partially. Lastly, only cyber-crime is defined by EWI or NICCS, satisfying guideline 4 [100-102]. We summarize our conclusions in Table 3 below. From this we can see that *cybercrime* meets the most guidelines of the conflicting terms. Clearly, cyber-crime is a term in great need of standardization, given the varied uses of its forms and synonyms. However, none of these three terms satisfies enough of our guidelines for us to recommend.

| Table 3 | Guidelines satisfied | | | |
|---|---|---|---|---|
| Term | 1 | 2 | 3 | 4 |
| Computer crime | O | X | ? | X |
| Cybercrime | X | O | O | X |
| Cyber crime | ? | ? | ? | O |

**Table 3. Forms of cyber-crime: X means the guideline is not satisfied, O means it is satisfied, ? means it is partially satisfied or not explicitly failed.**

## VIII. TOWARDS AN AGENDA AND METHODOLOGY

We argue for the creation of a cyber security research agenda for integrating all four categories into a unified cyber security discipline. We claim that such an agenda should be based on quantitative metadata from a repre-sentative sample of journal and conference papers. Below in Table 4 we categorize the terms from Table 2, and include the percentage of papers from each category that use Accepted and Partially Accepted terms, respectively, as well which papers use both, and citations.

| Proposed vocabulary for standardization, by category | | |
|---|---|---|
| Category | Accepted (A) | Partially Accepted (B) |
| Public (48.4%, 45.2%) (22.6% both) | cyber crime, cyber operations, cyber security, cyber warfare, cyberspace, DDOS, espionage, internet, national security, Stuxnet | Attack, CNI, Cybercrime, Cybersecurity, evidentiary, hacktivist/hactivist, information, missile defense, self-defense, signals intelligence, threat environment, u.s. cyber command, web |
| Infrastructure (60.0%, 80.0%) (60.0% both) | critical infrastructure, cyber physical systems, cyber security, DDOS, digital signature, network security, privacy, scada, steganography | Accountability, Availability, cascading failure, context-awareness, cps, industrial networks, risk assessment, information, Security, security architecture, sensitivity analysis, situational awareness, smart grid |
| Private (54.5%, 72.7%) (36.4% both) | CISO, cloud computing, computer abuse, cyber crime, cyber law, cyber security, cyberspace, internet, privacy, risk management, scada | Attack, cio, computer crime, cyber insurance, Cybersecurity, e-commerce law, information, insider, risks, Security, supply chain, supply chain management, threat patterns |
| General (68.4%, 89.5%) (63.2% both) | cyber crime, cyber law, cyber security, cyber threat, cyberspace, DDOS, denial of service, ICT, IDS, information technology, internet, intrusion detection system, malware, national security, phishing, privacy, system security | Attack, big data, Common vulnerabilities and exposures, computer ethics, computer system security, cps, cyber stalking, cybercrime, cybersecurity, dark web, forensics, information, security, semantic web, social-networking, software piracy, threat, vulnerability analysis, web |

**Table 4. Terms from Table 2, extracted from the literature review papers, and the categories of papers they appeared in. Some words appear in more than one category. Percentages indicate the percentage of papers that had at least one Accepted or Partially Accepted Term in a given category, respectively, and which had both. Average citations for papers using accepted and partially accepted, only accepted, only partially accepted, and not accepted terminology are given, labeled respectively as A+P, A, P, and N.**

Accepted and Partially Accepted keywords are roughly equally distributed across categories, meaning that every category individually does use a fair amount of accepted terminology. In addition, Infrastructure is the only one that doesn't mention cyber-crime, internet, or cyberspace. National security is shared between public and system. Cyber law shows up in private and system, but not public, curiously. More "application-driven or practical" concepts like malware, intrusion detection systems, forensics, big data, etc. appear only in System. Cyber security shows up in all four categories, as expected, and information shows up everywhere as well. DDOS is common to Public and Infrastructure. Privacy shows up in all categories except Public, strangely enough, and security as a standalone word shows up consistently in all other categories, yet only appears in one paper from Public. Attack is not a term used in any papers in the Private category. Cybersecurity [sic] as a single word shows up everywhere but Infrastructure, perhaps indicating that that spelling is less common in computer science. Infrastructure, on the other hand, is the only category with "smart grid."

Public has military terms like cyber operations and espionage, as well as national security terms like CNI and Stuxnet. Infrastructure has technical cyber security terminology like cyber physical systems, digital signature, accountability, and so forth; Private has many business aspects like cloud computing, CISO, cyber insurance, and computer abuse. General has a wide variety of general-sounding concepts: information technology, computer ethics, dark web, social-networking, and cyber threat.

In general, as expected, papers that use accepted terminology are lacking, but not scarce; and papers typically use more partially accepted terminology than accepted terminology, with some papers using both. Papers in the General category use more of both kinds, which is consistent with the definition of this category – they should use more accepted terminology because they are expected to be understood by a larger audience. Articles aimed at the public sector actually use the most non-standard terminology, which further calls into question why governments have been the sole authors of prior cyber security glossaries.

The sample size of this study is certainly too small to draw scientific conclusions. However, Table 4 provides another illustration of the four categories we derived from the literature review, and adds support to the notion of their existence. In addition, Table 4 gives some very real evidence of differences in communication between these four areas. If a larger sample size were taken, of 1000 papers or more – perhaps 10,000 – complete with distributions of which categories terms more commonly show up in as keywords or title words, it could be used in the formation of a research agenda for improving interdisciplinary cyber security research.

## IX. BROAD NOMENCLATURE ISSUES

While the previous discussion revolved around keywords extracted from a literature review of current trends in cyber security, it is by no means a comprehensive analysis, nor can such an analysis be done in a single paper. For this reason, in this section we slightly expand our analysis. Attention in the literature has recently been called to the variety of terms used to describe Internet-related concepts, with a number of different prefixes emerging since the Internet's creation, and achieving fluctuating levels of dominance over the years [115,116].

### A. INTERNET-RELATED PREFIXES

These words include *virtual*, *digital*, *e-*, *cyber*, *smart*, *net*, and *online*. If a proper systematic nomenclature is eventually to be constructed for researchers, the distinction, if any, between these words, should be understood, and redundant prefixes eliminated. The below descriptions refer to these words when used as prefixes or modifiers in computing.

#### 1) VIRTUAL

*Virtual* refers to that which seems real but isn't: simulation. This statement uses a loose definition of real, of course. In optics, *virtual* images are a phenomenon that results in the appearance of an image where no photons are actually present, i.e. that which seems real, but is not [124]. Virtual machines act like real ones but aren't. In fact, "virtual reality" could possibly refer to anything virtual (though obviously it conventionally refers to the human immersion in a virtual world). *Virtual* is typically for big picture things whose purpose is high-level. A virtual machine is not made to examine electrical signaling in computing, but to be operated by a user at the high level, for various purposes. Likewise, a *virtual* meeting room cares not about *how* the meeting takes place; it cares about the *contents* of the meeting, and simulating a meeting. This is of course the essence of high versus low levels of abstraction: low cares about how, high cares only about what.

#### 2) DIGITAL

*Digital* refers to something real, where the majority of the purpose of its being *digital* operates at a low level that is not visible, or which is a broad concept and not something humans can see right in front of them the way as they can with *virtual* things. Again, here, we use a loose interpretation of "real," and in fact, because of this loose definition, something can feasibly be both digital and virtual. For example, currency or the pixels of an image may be implemented at a low, bit level, i.e. digitally. Digital refers specifically to the digits involved – the bits of a computer. Digital processing of currency, images, and so forth, concerns itself with precisely *how* low level operations are performed. It is how pixels are programmed and represented, *in reality*, which makes an image digital.

Bitcoin, arguably a digital-virtual-currency, and is very concerned with the cryptographic algorithms involved in "mining" bitcoins. Bitcoin's status is, however, as of this writing, still controversial, so it is unclear how one should classify it. Digital currencies can typically be transformed between computers and a physical form, whereas virtual currencies cannot [125].

Digital currency and virtual meeting rooms are largely useless without the Internet. However, *digital* and *virtual* are not unique to networked technology. *Virtual* machines and *digital* images have no need for the Internet in order to function. Therefore, *virtual* and *digital* may more broadly be considered general "cyber" prefixes rather than Internets-specific prefixes. Furthermore, there is a clear distinction to be made between *virtual* and *digital* when used correctly. These two terms are primarily applied to words they modify to distinguish from "regular" versions of the words – e.g., a virtual machine, as opposed to a regular machine. Because of this, they do not directly contrast with each other, and can sometimes be used interchangeably.

#### 3) E-

*E-* means electronic, and refers to people-centric concepts like email, e-commerce, and e-residency. *E-* thus carries a distinct Internet and "popular accessibility" air with it. It is very much a 21st century term. If any of the prefixes in this section synonymous with *Internet*, it is *e-* or *net*. E-services or electronic services do not require the Internet to operate, though, but they generally do require some kind of network functionality. The IETF requires request for comments (RFC) documents spell email lowercase with no hyphen [126].

#### 4) CYBER

*Cyber* of course has its history in *cybernetics*, meaning *skilled in steering or governing*, and saw popular adoption and subsequent "official" usage by government and industry. It is a primary focus of this paper and needs no further introduction. *Cyber* is very much an Internet-age term, although it is not an exact synonym for *Internet*, but typically much broader in scope. We will not revise the definition of *cyber* here, since many other papers already define it – although none of the preeminent glossa-ries we mentioned earlier does [99-102,117,120]. Curiously, while "cybersecurity" [sic] saw large adoption as a security term in reference to computers, other terms (pre)modified by *cyber* have begun to emerge so quickly in recent years that they seem not to refer to "cyber" equivalents or corresponding aspects of real world phenomena, but to such phenomena as aspects of cyberspace; that is, "cyber" has become somewhat more of a possessive term and a noun adjunct rather than a modifying adjective. We believe that rather than, say, the cyber (aspect) of security, authors now speak of the security of cyber(space), perhaps unknowingly. Lastly, many authors claim that we have passed the "digital" age and are entering the cyber age [81].

### 5) SMART

*Smart* is a buzzword that emerged slowly in the 1990s as a reference to technology before taking off into mainstream vocabulary in the 2000s and skyrocketing in use in the early 2010s.[1] We predict that usage of smart will diminish in the coming age of the Internet of Things, since eventually appending "smart" to something will be superfluous – people may say, "well of course it's smart! It's electronic!" when discussing a modifier like this in the future. Therefore, we recommend it be used with caution and with the knowledge that it may be as obsolete in 10 years as many terms in the cyber security glossaries of 10 years ago are today.

### 6) NET

Lastly, "net," used as an adjunct noun when modifying another noun, refers explicitly to the Internet or sometimes another network, as a noun, rather than an adjective like *e-* does. It is thus the nominal synonym of Internet, whereas *e-* is the adjectival synonym. *Net*, like *e-*, has a narrow use than *cyber*. Unlike with *cyber*, which is ambiguously a noun or an adjective, in English it does not matter, in principle if net, as an adjunct noun, forms compounds as one or two words, though in practice *net* typically forms single-word compound nouns, such as netizens, NETmundial, and Netscape.

### 7) ONLINE

*Online* and *e* perform exactly the same function, but *e* is always a prefix (perhaps hyphenated) in a single-word compound, whereas *online* is a separate modifier.

### 8) INFORMATION TECHNOLOGY

So far, our discussions have not included information technology (IT) or information and communications technology (ICT), except that Table 4 shows them as accepted terms. Russia, for instance, sometimes considers information security, rather than cyber security; and IT has a different connotation than *cyber* [129]. The ITU heavily promotes usage of ICT, and IT/ICT security is sometimes viewed as a subset of cyber security focusing only on information and no other concerns – nevertheless, the exact definition of ICT is generally highly contested [130-131]. Elsewhere, IT is seen as a physical substrate for cyberspace. In addition, cyber is a more flexible English modifier than IT or ICT. In our opinion, IT and ICT are unstable terms and, where possible, *cyber* should be used instead. It is important to maintain clear and consistent language to facilitate knowledge sharing across disciplines.

### B. A UNIFYING ACADEMIC DISCIPLINE NAME

While cyberspace is becoming an increasing security concern, it is also becoming ubiquitous as an aspect of the human experience, which is becoming less separable every year from issues cyberspace combines. Social engineering is a prime example of cyberspace and in particular, cyber security, bleeding into the human psychological realm. It is equally important for scholars to unite in research surrounding this general "cyber" field, just as they should with security. This conjugation of cyberspace and physical space, and the constant growth of new *cyber* terminology, ad hoc or not, is leading to the formation of a new academic discipline, a so-called *Cybermatics* field according to Ma et al. (defined below), with emphasis on creating new terms to describe characteristics of *cyberspace* such as "cyber-something" in either *real* or virtual terms, rather than seeking to describe characteristics of the *real world* in terms of computers and *cyberspace* (such as security, adapted for cyberspace: *cybersecurity* [sic]), as was done in the early years of the Internet. This influx of terms warrants closer inspection and regulation, lest valuable knowledge generated by scholars go unnoticed by researchers unfamiliar with these ad hoc terms; this is a potential problem when searching journal databases without knowing the right keywords to search for, as stated earlier.

---

[1] http://www.scopus.com/term/analyzer.url?sid=7148782FEA989C1354BD1E385A58EF9B.I0QkgbIjGaqLQ4Nw7dqZ4A%3a60&origin=resultslist&src=s&s=%28TITLE-ABS-KEY%28smart%29+AND+TITLE-ABS-KEY%28computer%29OR+TITLE-ABS-KEY%28internet%29%29&sort=plf-f&sdt=b&sot=b&sl=76&count=31202&analyzeResults=Analyze+results&txGid=0

Although the term *cyber* is being used more and more frequently, it is used in a variety of contexts, both technical and nontechnical in nature. This domain of research and knowledge extends beyond cyber security and includes general issues of Internet governance and online behavior. Recently, Ma et al. proposed the term "Cybermatics" to describe this new field that encompasses all things cyber and cyber-related [81]. This includes both concepts within cyberspace (Ma et al.'s so called "Cyber World"), such as cyberbullying, and concepts of utilization of cyberspace ("cyber-conjugated" or "cyberization"), such as cyber-physical systems.

In their paper, Ma et al. first define cyber entities, as "anything that exists digitally in cyberspace, either purely synthesized by a computer, or closely correlated and further conjugated with a real entity in physical, social and mental spaces" [81]. They go on to define Cybermatics as a holistic field which studies cyber entities and their properties, models, and representations, including their relations and conjugations, and their technologies and applications.

Although the intention of this paper was to search for cyber security journal papers, many conclusions drawn from it are shared throughout Cybermatics. We now briefly linguistically analyze whether Cybermatics is an appropriate name for the "Cyber" knowledge domain, and propose alternative labels.

## 1) ETYMOLOGY OF CYBERMATICS
We believe it is necessary to standardize a term to unify the academic study of cyber-related concepts. Ma et al. (2015) give the etymology of their proposed term "Cybermatics" for the new "cyber" field:

'The suffix *-matic* comes from *matos* in Greek that means "willing to (perform)". The suffix *-ic* comes from *-ikos* in Greek, meaning "behaving like" or "having the characteristics of". The suffix *-ics* can be used to form a noun to name a field of study, for instance, mathematics, automatics, kinematics, systematics, and so forth. The term "*cybermatic*" can be regarded as "cyber + *matos* + *ikos*", which may describe a thing willing/able to be, behaving like or having cyber characteristics. In a linguistic sense, "Cybermatics" can be understood as a field in which cybermatic things, i.e., various cyber entities existing in cyber-enabled worlds as distinct phenomena, are studied' [81].

Given Ma et al.'s description of Cybermatics throughout their paper, we think that Cybermatics as an overarching field for all things cyber – whether in the "Cyber World" or whether they are "Cyber-conjugated" – is possible. However, the name "Cybermatics" is unlikely to be widely accepted, and at this stage it is too early to predict adoption. To facilitate the adoption of an overarching term, we believe it is helpful for the academic community to choose from a number of candidate terms. While "cyber" has its basis in computer science, its transdisciplinary nature necessitates input from many bodies. Therefore, the academic community referenced here should consist of all parties with a stake in this field.

## 2) ALTERNATIVE ACADEMIC DISCIPLINE NAMES
We now suggest potential alternative transdisciplinary field names, for consideration by scholars. These suggestions are meant only as possibilities, and we hope that if any of these terms is adopted, only one is. However, we feel that considering multiple terms for adoption is the best way to determine the most appropriate one for standardization.

An examination of a large number of academic disciplines revealed some of the following suffixes: *-matics*, *-ology*, *-nomics* or *-nomy*, science, *-ry*, *-ic*, *-istics*, *-ation*, studies, and *-graphy* [108]. Of these suffixes, three stand out: "Cyber science", "Cyberistics", and "Cybernomics". "Cyber science" ironically does not have the futuristic feeling of the other two (or Cybermatics), and its etymology requires little exploration. We do however propose it as a possible field name. It should be noted, however, that Ma et al. propose Cyber Science as only one subdiscipline of Cybermatics. For "cyberistics," *-istics* is made from two suffixes, *-ism* and *–ic,* and the latter is used is Cybermatics and is etymologically sensible. However, *-ism* refers to a doctrine, practice, or system, and derives from Greek *-ismos*, meaning the practice or teaching of a thing. [109] "Cyber" is not a practice or doctrine, so this suffix is not appropriate. Of the above three candidates, "cybernomics" is the most interesting (pronounced like genomics). While genomics derives from a neologism "-omics," which has specifically biological applications, the root of economics refers to law, custom, rule, ordinance, or management [110, 111]. One might speak of the laws governing cyberspace (artificial or natural), or what might speak of the entirety of activities related to cyberspace, as the biological –ominics can carry the sense of "all constituents considered collectively."

We therefore propose "cybernomics" as a reasonable candidate term encompassing the "cyber" academic discipline, in competition with "Cybermatics", "Cyber science", and indeed, perhaps the frontrunner candidate, "Cyber". In our opinion, *cyber* is likely to emerge the winner among these terms because of its prevalence, but we do not advocate adoption of any particular term herein. We do, however, advocate adopting a standard term for the field in the near future, by official standards bodies, governing bodies, research institutions, and governments, just as we propose an updated cyber security dictionary.

## X. CONCLUSION

Many authors still use ad hoc terms despite the existence of standards glossaries, and spelling or phrasing of many terms is still not agreed upon. The lack of collaboration across disciplines inferred from our review emphasizes the need for more comprehensive standard terminology for both cyber security and broader cyber research. Except when radically new concepts are written about, greater use of more widely accepted terms is recommended, though not at the expense of innovation. Authors should, before submitting for publishing, search the databases for their potential keywords to ensure that all are in the [10,100000) range, and that at least one is in the [100,1000) range to ensure good searchability. Because the papers reviewed were necessarily all recently published, and not all from the same year, (2010-2015), it is difficult to determine any correlation between type of vocabulary used and citations. Future research could aim to verify whether such a correlation exists – a positive one could bolster efforts toward adoption of standard vocabulary. However, we believe that regardless, there are compelling reasons to update existing cyber security glossaries.

We outlined guidelines to use when considering keywords to use in future publications and when crafting terminology standards, and resolved some long-held misconceptions in spelling and phrasing. We encourage use of these guidelines and the following recommendations, as well as the use of the standard glossary projects from EWI, NICCS, and other complementary sources like NISTIR 7298. These existing dictionaries are, however, mostly constructed by the public sector, and may or may not reflect academic and private sector areas of study and work regarding cyber security. Therefore, greater effort from outside of governments, and collaboration with the greater global multistakeholder community, is essential when creating or updating cyber security glossaries.

We proposed a classification of research areas concerned with cyber security, which can be refined by a more comprehensive study of keywords comprising it. These keywords can be used to craft research agendas for each area, as well as in crafting cross-disciplinary research agendas for cyber security. Within the categories we identified, use of standard terminology is fairly common. However, there is clear room for improvement among authors and working groups. Other possible categorizations may consist of the common social sectors of civil society, industry, academia, and the government that many articles cite [127]. We encourage future researchers to delve further into categorization and ontology creation of cyber security for the formulation of research agendas.

Specific spelling and phrasing conventions should be adhered to in order to ensure visibility of publications. Most importantly, except in the cases of cyberspace, "cyber" terms should be written with cyber as a separate word, as in "cyber physical," possibly hyphenated. While cyber security is the prevailing spelling, it is reasonable to assume that the single word spelling, cybersecurity, is still acceptable. Cyber-crime has no definitive spelling, but we predict it will lean toward being condensed to cybercrime in the future.

Herein we attempted to lay the groundwork for standardizing communication within cyber security, in order to begin to formalize the scientific methodology of the field. We believe formalizing cyber security would accelerate the pace of research, improve policymaking and business practice, and lead to greater integration with the rest of the scientific community. Addition efforts that may be important to formalizing cyber security as an academic discipline include the creation of more businesses out of research, the creation of a committee within an internet governance body, or the formation of a multistakeholder project, to address this, and systematic efforts by academics to propose, assess, and rigorously define vocabulary based on the 4 guidelines given in this paper. The ultimate goal of such formalization should not be simply a lexicon of terminology, but methodologies or framework for cyber security research. With the growing prevalence of cyberspace and the emergence of a so-called Cyber or Cybernomics or Cybermatics field, it is urgent to bring together the disparate efforts in these areas and share knowledge, lest it be overlooked and progress delayed.

**This work was funded by the Cooperative Agreement between the Masdar Institute of Science and Technology (Masdar Institute), Abu Dhabi, UAE and the Massachusetts Institute of Technology (MIT), Cambridge, MA, USA - Reference 02/MI/MIT/CP/11/07633/GEN/G/00**

# XI. APPENDIX

## Figure A1.

Logarithmically-scaled Scopus, IEEE Xplore, and combined search incidences of keywords extracted from reviewed articles, arranged in alphabetical order.

Figure A1a displays the y-axis values, the horizontal lines of which are carried through to the other figures, and the incidences of "academia" through "cryptography"

Figure A1b shows incidences of "cryptology" through "cyber-territory"

Figure A1c shows incidences of "dark web" through "information systems security"

Figure A1d shows incidences of "information technology" through "ontology-based context models"

Figure A1e shows incidences of "organizational justice" through "semantic web technology"

Figure A1f shows incidences of "semantic web technology" through "web space"

**Figure A1a.** Figure A1a displays the y-axis values, the horizontal lines of which are carried through to the other figures, and the incidences of "academia" through "cryptography"

**Figure A1b.** Scopus, IEEE Xplore, and total Incidences of "cryptology" through "cyber-territory"

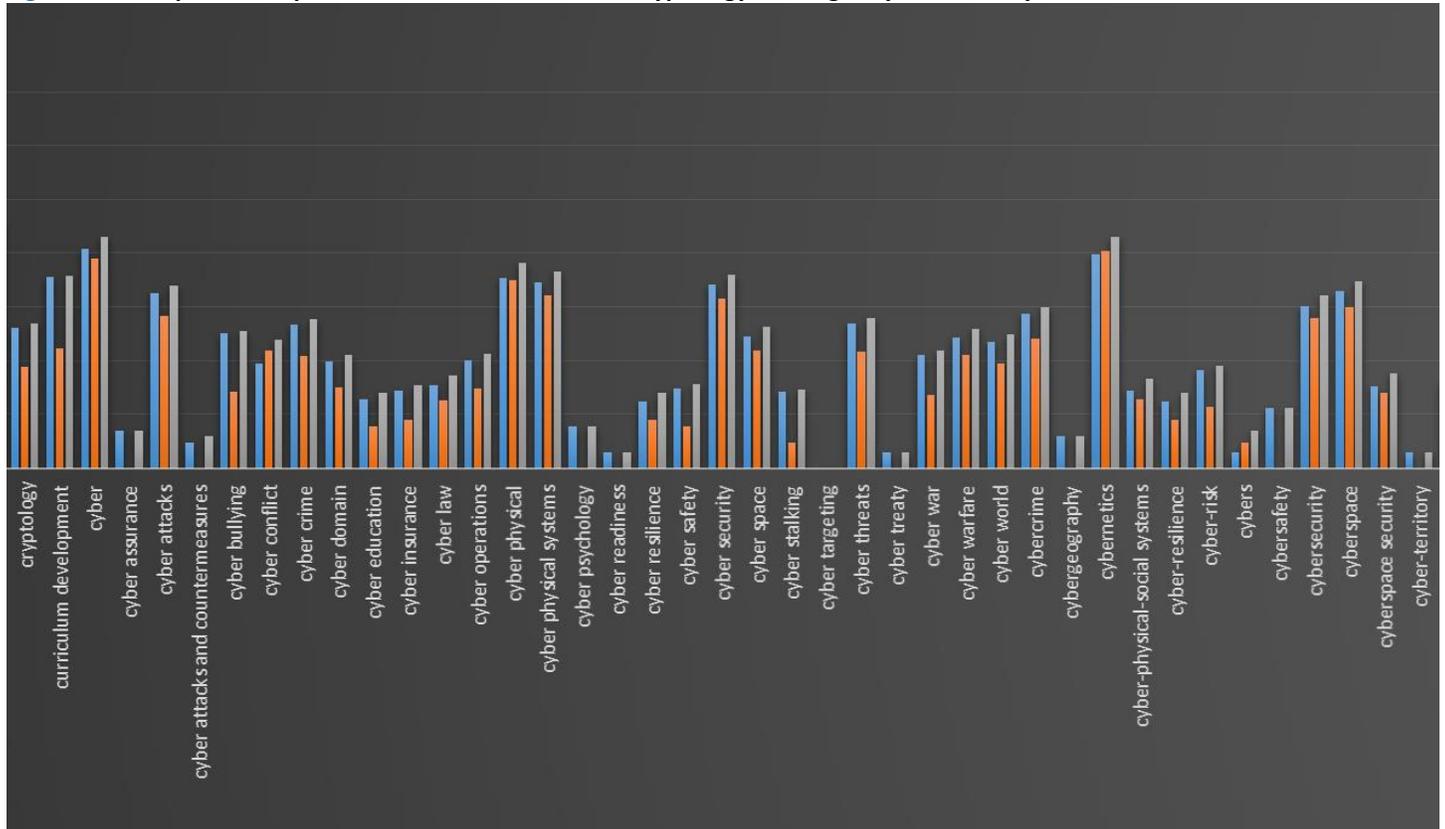

**Figure A1c.** Scopus, IEEE Xplore, and total Incidences of "dark web" through "information systems security"

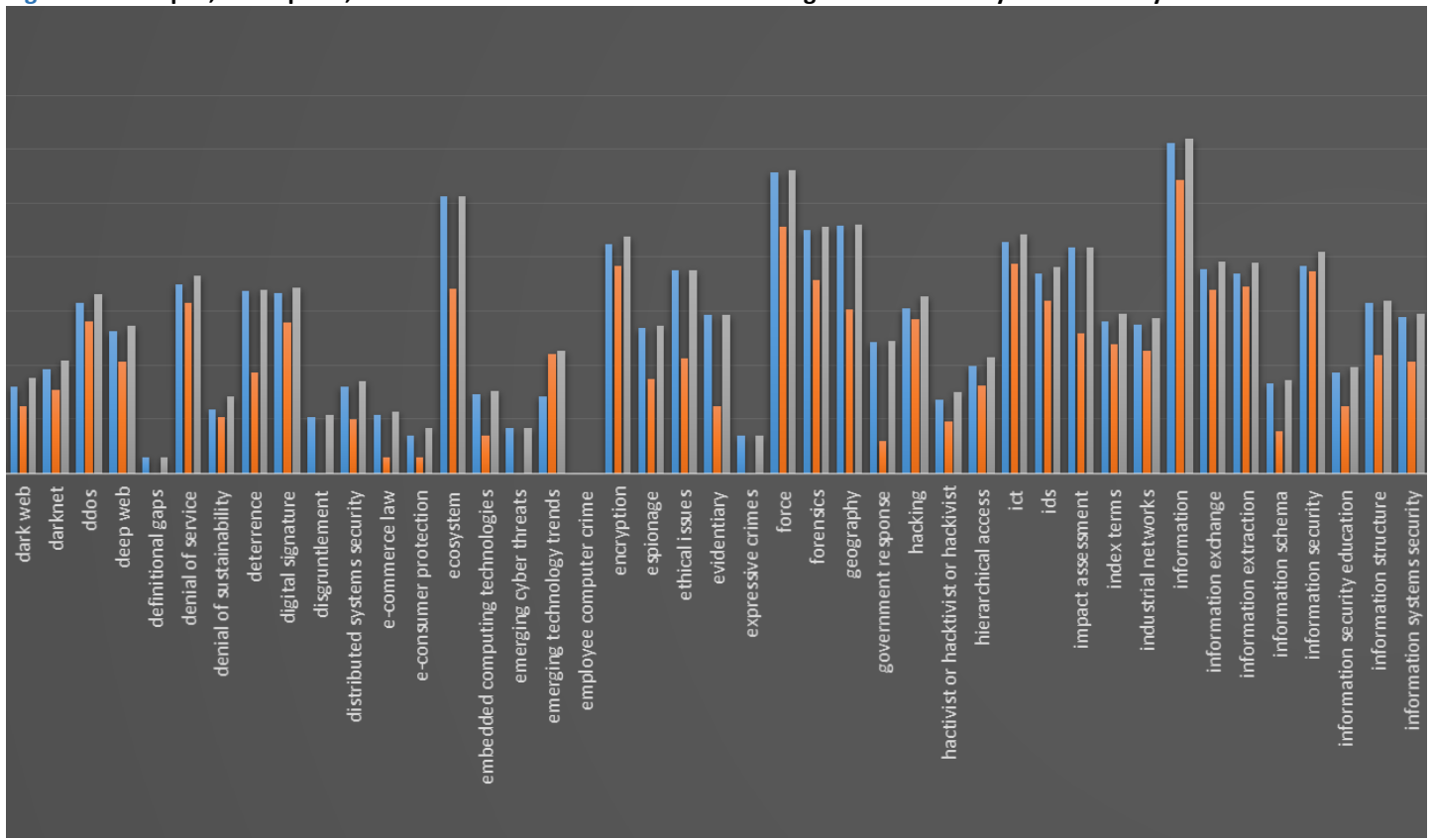

**Figure A1d.** Scopus, IEEE Xplore, and total Incidences of "information technology" through "ontology-based context models"

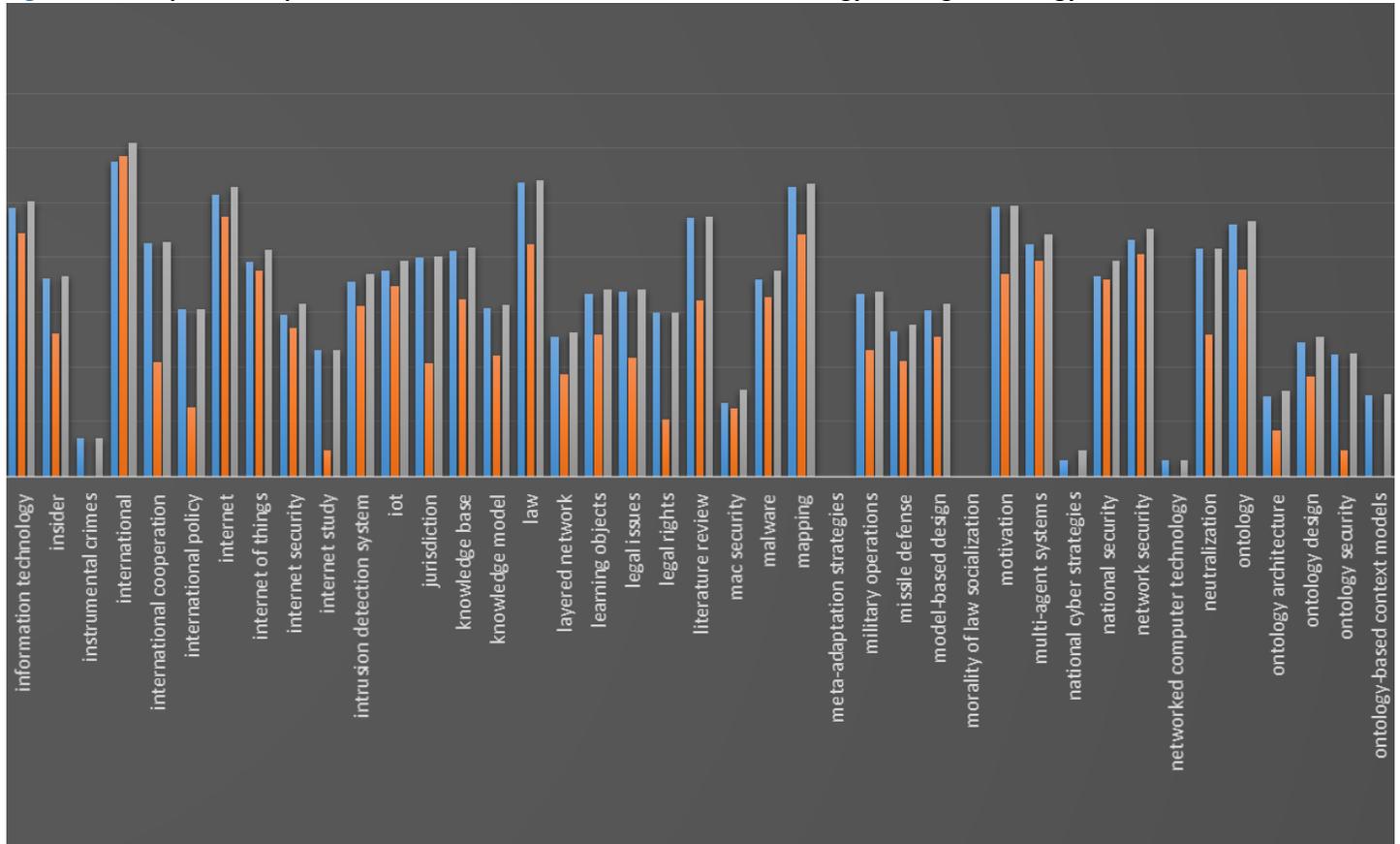

**Figure A1e.** Scopus, IEEE Xplore, and total Incidences of "organizational justice" through "semantic web technology"

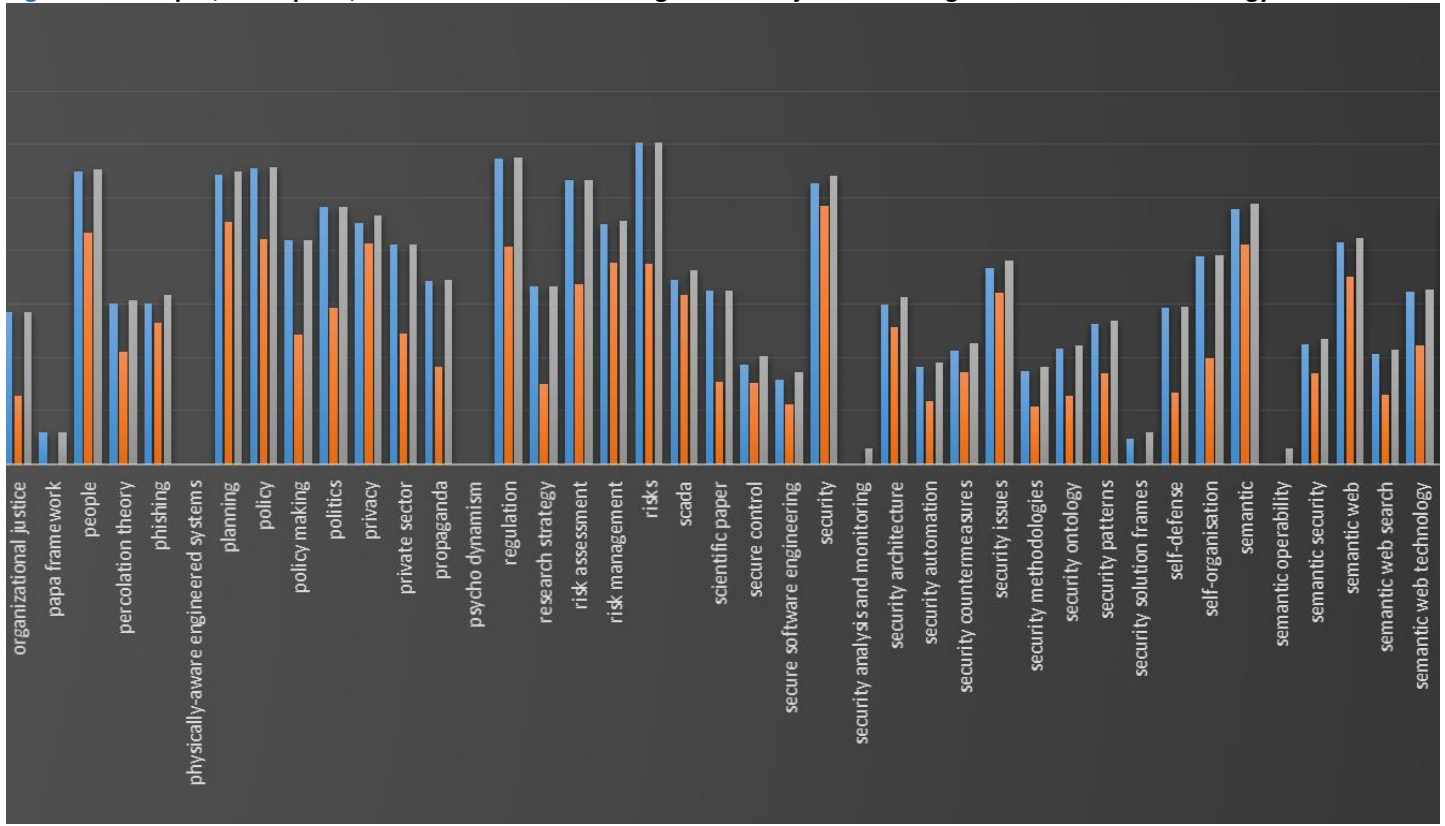

**Figure A1f** Scopus, IEEE Xplore, and total Incidences of "semantic web technology" through "web space"

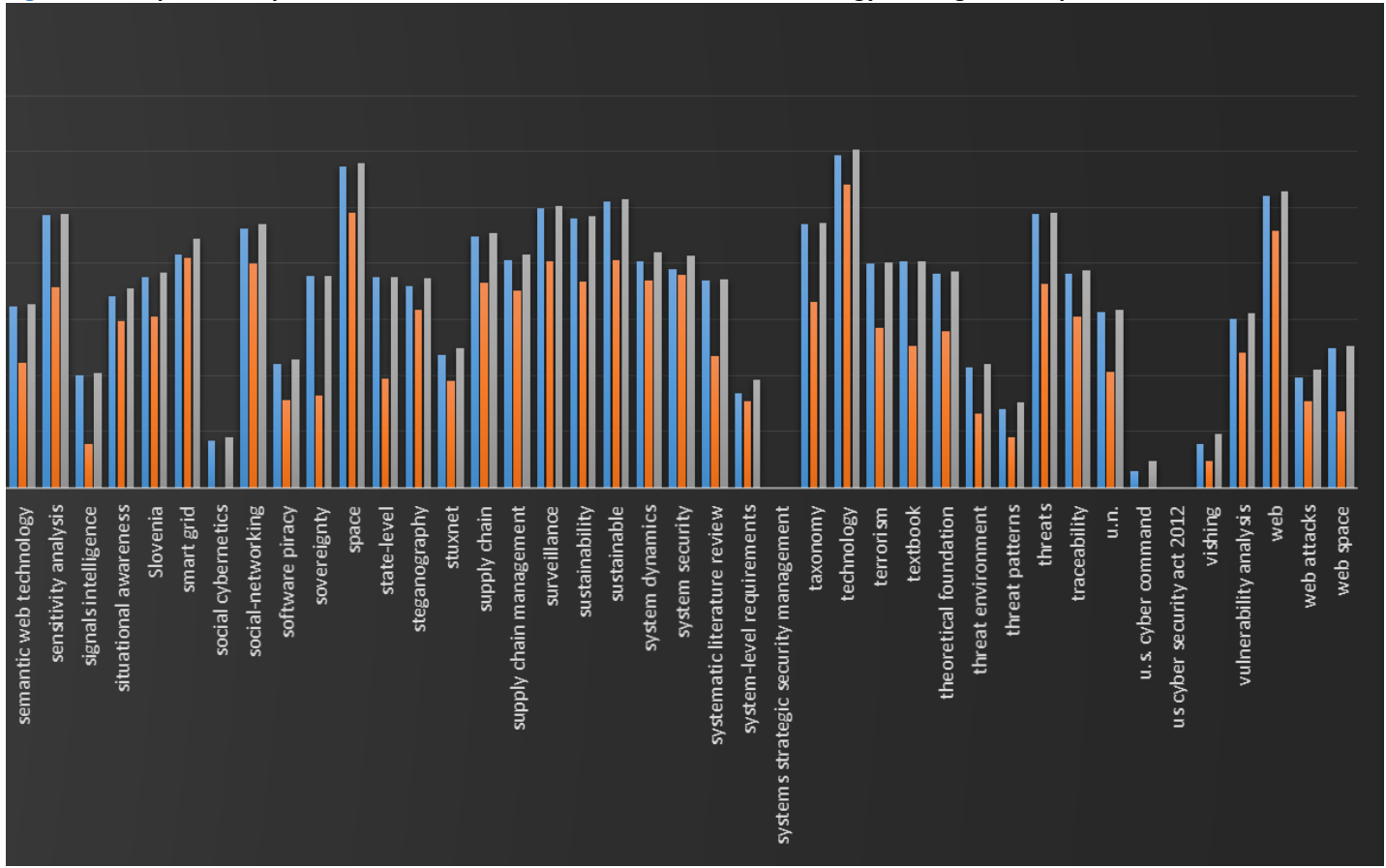

**Table A1.**

Keywords extracted from literature review, sorted by powers of 10 of the number of results returned by searching Scopus and IEEE Xplore

| Category (number of hits) | Term (Scopus) | | Term (IEEE Xplore) | | Term (Total) | |
|---|---|---|---|---|---|---|
| 1 (1,000,000+) | information | risks | | | information | risks |
| | | | | | international | technology |
| 2 [100000,1000000) | attack | people | communication | international | attack | people |
| | availability | planning | information | technology | availability | planning |
| | classification | policy | | | classification | policy |
| | communication | regulation | | | communication | regulation |
| | ecosystem | risk assessment | | | ecosystem | risk assessment |
| | force | security | | | force | security |
| | | | | | information | |
| | international | space | | | technology | space |
| | internet | sustainable | | | internet | surveillance |
| | law | technology | | | law | sustainable |
| | mapping | web | | | mapping | web |
| 3 [10000,100000) | academia | network security | attack | people | academia | network security |
| | accountability | neutralization | availability | planning | accountability | neutralization |
| | attribution | ontology | classification | policy | attribution | ontology |
| | big data | policy making | cloud computing | privacy | big data | policy making |
| | cloud | | | | | |
| | computing | politics | cybernetics | regulation | cloud computing | politics |
| | complex | | | | | |
| | networks | privacy | force | security | complex networks | privacy |
| | | | information | | | private |
| | cryptography | private sector | technology | semantic | cryptography | sector |
| | cyber | risk management | internet | smart grid | cyber | risk management |
| | encryption | semantic | law | space | cybernetics | semantic |
| | forensics | semantic web | mapping | surveillance | encryption | semantic web |
| | geography | sensitivity analysis | network security | sustainable | forensics | sensitivity analysis |
| | ict | smart grid | | web | geography | smart grid |
| | impact assessment | social-networking | | | ict | social-networking |
| | information technology | supply chain | | | impact assessment | supply chain |
| | international co-operation | supply chain management | | | information security | supply chain management |
| | jurisdiction | surveillance | | | international cooperation | sustainability |
| | knowledge base | sustainability | | | internet of things | system dynamics |
| | literature review | system dynamics | | | jurisdiction | security |
| | motivation | taxonomy | | | knowledge base | taxonomy |
| | multi-agent systems | textbook | | | literature review | terrorism |
| | | threats | | | motivation | textbook |
| | | intrusion detection system | | | | |
| 4 [1000,10000] | ami | detection system | academia | knowledge base | ami | iot |
| | computer security | iot | | literature review | cascading failure | knowledge model |
| | context-awareness | knowledge model | big data | review | cio | learning objects |
| | cps | learning objects | complex networks | malware | computer security | objects |
| | cpss | legal issues | computer security | motivation | context-awareness | legal issues |
| | critical infrastructures | malware | context-awareness | multi-agent systems | cps | legal rights |
| | | military | cps | national security | cpss | malware |
| | cryptanalysis | operations | cryptography | ontology | critical infrastructures | military |
| | curriculum development | model-based design | cyber | risk assessment | cryptanalysis | operations |
| | cyber attacks | national security | cyber physical | risk management | curriculum development | model-based design |
| | | percolation theory | cyber physical systems | risks | cyber attacks | national security |
| | cyber physical | | cyber security | scada | cyber physical | percolation theory |
| | cyber physical systems | phishing | denial of service | security issues | cyber physical systems | |
| | cyber security | propaganda | ecosystem | semantic web | cyber security | phishing |
| | cybernetics | research strategy | encryption | sensitivity analysis | cybersecurity | propaganda |
| | cybersecurity | scada | forensics | Slovenia | cyberspace | research strategy |
| | cyberspace | scientific paper | geography | social-networking | ddos | scada |
| | ddos | security issues | ict | steganography | denial of service | scientific paper |
| | denial of service | self-organisation | ids | supply chain | deterrence | security architecture |
| | | semantic web technology | information exchange | supply chain management | digital signature | security issues |
| | deterrence | situational awareness | information extraction | sustainability | ethical issues | self-organisation |
| | digital signature | Slovenia | information security | system dynamics | hacking | semantic web technology |
| | ethical issues | | internet of things | security | ids | situational awareness |
| | hacking | sovereignty | intrusion detection system | taxonomy | information exchange | Slovenia |
| | ids | state-level | iot | threats | information extraction | sovereignty |
| | information exchange | steganography | | traceability | information structure | state-level |
| | information extraction | system security | | | insider | steganography |
| | information security | systematic literature review | | | international policy | systematic literature review |
| | information structure | terrorism | | | internet security | theoretical foundation |
| | | theoretical foundation | | | intrusion detection system | traceability |
| | insider | | | | | u.n. |
| | international policy | traceability | | | | vulnerability analysis |
| | internet of things | u.n. | | | | |
| | | vulnerability analysis | | | | |
| 5 [100,1000] | anti-forensics | industrial networks | accountability | information structure | anti-forensics | hierarchical access |
| | | information systems security | | information systems security | ciso | index terms |
| | cascading failure | security | ami | security | | |

| | | | | | |
|---|---|---|---|---|---|
| cio | internet security | attribution | insider | cni | Industrial networks information systems security |
| cni | internet study | cascading failure | international cooperation | computer ethics | |
| computer ethics | layered network | cio | internet security | computer system security | internet study |
| computer system security | legal rights | cpss | jurisdiction | cryptology | layered network |
| cryptology | missile defense | critical infrastructures | knowledge model | cyber bullying | missile defense |
| cyber bullying | ontology design | cryptanalysis | learning objects | cyber conflict | ontology design |
| cyber crime | ontology security | curriculum development | legal issues | cyber crime | ontology security |
| cyber operations | organizational justice | cyber attacks | military operations | cyber domain | organizational justice |
| cyber space | security architecture | cyber conflict | missile defense | cyber operations | secure control |
| cyber threats | security countermeasures | cyber crime | model-based design | cyber space | security countermeasures |
| cyber war | security ontology | cyber space | neutralization | cyber threats | security ontology |
| cyber warfare | security patterns | cyber threats | percolation theory | cyber war | security patterns |
| cyber world | self-defense | cyber warfare | phishing | cyber warfare | self-defense |
| cybercrime | semantic security | cybercrime | policy making | cyber world | semantic security |
| deep web | semantic web search | cybersecurity | politics | cybercrime | semantic web search |
| espionage | signals intelligence | cyberspace | private sector | darknet | signals intelligence |
| evidentiary government response | software piracy | ddos | security architecture | deep web | software piracy |
| | stuxnet | deep web | semantic web | emerging technology trends | |
| index terms | threat environment | digital signature | situational awareness | espionage | stuxnet |
| | web space | emerging technology trends | systematic literature review | evidentiary government response | threat environment |
| | | ethical issues | terrorism | | web attacks |
| | | hacking | textbook | | web space |
| | | impact assessment | theoretical foundation | | |
| | | index terms | u.n. | | |
| | | industrial networks | vulnerability analysis | | |

| 6 [10,100) | | | | | | |
|---|---|---|---|---|---|---|
| | active cyber defense | denial of sustainability | anti-forensics | layered network | active cyber defense | dark web |
| | ciso | disgruntlement | cni | legal rights | cikr | denial of sustainability |
| | computational part | distributed systems security | computational part | mac security | computational part | disgruntlement distributed systems security |
| | computer abuse | e-commerce law embedded computing technologies | computer ethics | ontology design | computer abuse | |
| | critical national infrastructure | | computer system security | organizational justice | critical national infrastructure | e-commerce law embedded computing technologies |
| | | emerging technology trends | critical national infrastructure | propaganda | | |
| | cyber conflict | hacktivist/hacktivist | cryptography | research strategy | cyber education | hacktivist/hacktivist |
| | cyber domain | hierarchical access | cyber bullying | scientific paper | cyber insurance | information schema |
| | cyber education | information schema | cyber domain | secure control | cyber law | information security education |
| | cyber insurance | information security education | cyber law | secure software engineering | cyber resilience | |
| | cyber law | mac security | cyber operations | security automation | cyber safety | mac security |
| | cyber resilience | | cyber war | security countermeasures | cyber stalking | ontology architecture |
| | cyber safety | ontology architecture | cyber world | security methodologies | cyber-physical-social systems | ontology-based context models |
| | cyber stalking | ontology-based context models | cyber-physical-social systems | security ontology | cyber-resilience | secure software engineering |
| | cyber-physical-social systems | secure control | cyber-risk | security patterns | cyber-risk | security automation |
| | cyber-resilience | secure software engineering | | | cybersafety | security methodologies |
| | cyber-risk<br>cybersafety<br>cyberspace security<br>dark web<br>darknet | security automation<br>security methodologies<br>system-level requirements<br>threat patterns<br>web attacks | cyberspace security<br>dark web<br>darknet<br>denial of sustainability<br>deterrence<br>distributed systems security<br>espionage<br>evidentiary<br>hierarchical access<br>information security education<br>international policy | self-defense<br>self-organisation<br>semantic security<br>semantic web search<br>software piracy<br>sovereignty<br>state-level<br>stuxnet<br>system-level requirements<br>threat environment<br>web attacks<br>web space | cyberspace security | system-level requirements<br>threat patterns |
| 7 [0,10) | active air defense<br>adaptation tactics<br>attack description language<br>cikr<br>common vulnerability exposures<br>cross-domain attacks<br>cyber assurance<br>cyber attacks and countermeasures<br>cyber psychology<br>cyber readiness<br>cyber targeting<br>cyber treaty<br>cybergeography<br>cybers<br>cyber-territory<br>definitional gaps<br>e-consumer protection<br>emerging cyber threats | employee computer crime<br>expressive crimes<br>instrumental crimes<br>meta-adaptation strategies<br>morality of law socialization<br>national cyber strategies<br>networked computer technology<br>papa framework<br>physically-aware engineered systems<br>psycho dynamism<br>security analysis and monitoring<br>security solution frames<br>semantic operability<br>social cybernetics<br>systems strategic security management<br>u.s. cyber command<br>us cyber security act 2012<br>vishing | active air defense<br>active cyber defense<br>adaptation tactics<br>attack description language<br>cikr<br>ciso<br>common vulnerability exposures<br>computer abuse<br>cross-domain attacks<br>cyber assurance<br>cyber attacks and countermeasures<br>cyber education<br>cyber insurance<br>cyber psychology<br>cyber readiness<br>cyber resilience<br>cyber safety<br>cyber stalking<br>cyber targeting<br>cyber treaty<br>cybergeography<br>cyber-resilience<br>cybers<br>cybersafety<br>cyber-territory<br>definitional gaps<br>disgruntlement<br>e-commerce law<br>e-consumer protection | embedded computing technologies<br>emerging cyber threats<br>employee computer crime<br>expressive crimes<br>government response<br>hacktivist/hacktivist<br>information schema<br>instrumental crimes<br>internet study<br>morality of law socialization<br>national cyber strategies<br>networked computer technology<br>ontology architecture<br>ontology security<br>ontology-based context models<br>papa framework<br>physically-aware engineered systems<br>psycho dynamism<br>security<br>analysis and monitoring<br>Security<br>solution frames<br>semantic operability<br>signals intelligence<br>social cybernetics<br>systems strategic security management<br>threat patterns<br>u.s. cyber command<br>us cyber security act 2012<br>vishing | active air defense<br>adaptation tactics<br>attack description language<br>common vulnerability exposures<br>cross-domain attacks<br>cyber assurance<br>cyber attacks and countermeasures<br>cyber psychology<br>cyber readiness<br>cyber targeting<br>cyber treaty<br>cybergeography<br>cybers<br>cyber-territory<br>definitional gaps<br>e-consumer protection<br>emerging cyber threats | employee computer crime<br>expressive crimes<br>instrumental crimes<br>meta-adaptation strategies<br>morality of law socialization<br>national cyber strategies<br>networked computer technology<br>papa framework<br>physically-aware engineered systems<br>psycho dynamism<br>security analysis and monitoring<br>security solution frames<br>semantic operability<br>social cybernetics<br>systems strategic security management<br>u.s. cyber command<br>us cyber security act 2012<br>vishing |